\documentclass[12pt]{article}
\usepackage[utf8]{inputenc}
\DeclareUnicodeCharacter{2212}{-}
\usepackage{xcolor, graphicx, authblk}
\usepackage{amsfonts,amsmath,amssymb,mathtools}
\usepackage[left=2cm,right=2cm,top=2cm,bottom=2cm,bindingoffset=0cm]{geometry}
\usepackage{cite}
\usepackage{tabularx}
\usepackage{mathrsfs}
\usepackage{subcaption}
\usepackage[scr=boondox]  
           {mathalpha}
\usepackage[english]{babel}

\usepackage[T1]{fontenc}
\usepackage{esint}

\usepackage{soul}

\usepackage{setspace}
\usepackage{braket}
\usepackage[labelfont=bf, margin=20pt,font={small}]{caption}
\usepackage{subcaption}
\usepackage{youngtab}
\usepackage{epigraph}
\usepackage{upgreek}

\newcommand\beq{\begin{equation}}
\newcommand\eeq{\end{equation}}

\usepackage{import}
\usepackage{xifthen}
\usepackage{pdfpages}
\usepackage{transparent}

\graphicspath{{figures/}}

\definecolor{BlueGreen}{RGB}{49,152,255}
\definecolor{Violet}{RGB}{120,80,120}
\definecolor{Blue}{RGB}{0,0,255}
\definecolor{Yellow}{RGB}{0,255,51}
\definecolor{ElectricGreen}{RGB}{0, 255, 0}
\definecolor{MediumPersianBlue}{RGB}{0, 103, 165}
\definecolor{gray}{RGB}{0, 95, 95}

\usepackage[unicode, colorlinks, urlcolor=gray, linkcolor=gray, citecolor=Blue]{hyperref}
\numberwithin{equation}{section}



\title{\bf Effective graviton mass in de Sitter space}

\author[1,2]{D.~I.~Sadekov\thanks{\href{mailto:sadekov.di@phystech.edu}{sadekov.di@phystech.edu}}}
\affil[1]{Moscow Institute of Physics and Technology, 141701, Institutskiy pereulok 9, Dolgoprudny, Russia}
\affil[2]{\textcolor{black}{NRC ”Kurchatov Institute”, 123182, Moscow, Russia}}
\date{\today}

\begin{document}

\maketitle

\begin{abstract}
    We calculate the effective mass of gravitational perturbations induced by the interaction of the classical gravitational field with quantum matter in the background of the Poincaré patch of de Sitter space. Using the Schwinger-Keldysh diagrammatic technique, the one-loop effective action is calculated and it is shown that the graviton does not acquire mass for the most symmetric Bunch-Davies state. However, we have shown that even in this case, there is a nontrivial modification of the theory at one loop in the scalar sector of gravity.
\end{abstract}

\newpage

\tableofcontents

\newpage
\section{Introduction}\label{sec:Introduction}
Quantum field theory in curved space-time aims to shed light on the problems of the cosmological constant and the evolution of the early Universe. De Sitter space is the simplest example for investigating these questions, but there are still many subtleties that have not been studied so far in sufficient details, such as IR divergences in loop corrections \cite{Krotov:2010ma, Akhmedov:2017ooy, Akhmedov:2019cfd}, vacuum instabilities \cite{Anderson:2013ila, Anderson:2013zia, Krotov:2010ma, Akhmedov:2019esv}, and the behavior of light fields \cite{Akhmedov:2017ooy}. One way to explore the behavior of the system and the response of quantum matter to external conditions is to find the effective action for small perturbations of the external field. This paper’s main objective is to study the effective mass term for the graviton, which it may acquire in the one-loop effective action in de Sitter (dS) and anti-de Sitter (AdS) space-times. The motivation for this question comes from the natural Gibbons-Hawking temperature in dS \cite{Popov:2017xut, Birrell:1982ix}, which suggests that photon and graviton can acquire a non-zero mass as it happens in the physics of plasma. Our work is inspired by the paper \cite{Popov:2017xut}, where it is shown that despite the fact that an observer would detect some sort of thermal equilibrium with the canonical temperature $T_{\text{dS}} = \frac{H}{2\pi}$ ($H$ is the Hubble constant here), there is no effective Debye mass for photon for the most symmetric Bunch-Davies state of the matter. We extend this discussion onto the case of the gravitational mass. The graviton field itself is considered at the classical level as a perturbation of the dS metric, and we consider free scalar field theory as the quantum matter. It is worth noting that perturbation of the metric in an external field has several physical modes that can exhibit different behavior in the effective theory. Consideration of these cosmological perturbations is important for understanding the propagation of gravitational waves and density fluctuations of matter in the early Universe \cite{Gorbunov:1354521}.

There are two well-studied types of massive terms \cite{Fierz:1939ix, deRham:2014zqa} that can be added to the gravity action:
\beq \label{eq:massive_gravity}
    S_{\text{mass}} = \int d^Dx\sqrt{|g|}\left[\epsilon_{\text{gh}}h^2 + m_{\text{g}}^2\left(h^{\mu\nu}h_{\mu\nu} - h^2\right)\right],
\eeq
which break diffeomorphism invariance. The second one is called Fierz-Pauli massive term and it is known to bring no ghost-like degrees of freedom to the linearized gravity, while the first one is associated to the so called ``scalar ghost'' and leads to Ostrogradsky's instability \cite{deRham:2014zqa}. In our work, we attribute the emergence of mass to the appearance of terms like (\ref{eq:massive_gravity}), if any, in the long-wave expansion of the effective action, which in turn does respect gauge invariance. For instance, given the Minkowski background, we have for Ricci scalar $R$ in the linear and second orders:
\beq\label{eq:linearized_R_in_Mink}
\begin{aligned}
    R^{(1)} &\sim k^2\left(g_{\mu\nu} - \frac{k_{\mu}k_{\nu}}{k^2}\right)h^{\mu\nu},\\
    \left(\sqrt{|g|}R\right)^{(2)} &\sim k^2\left(h^{\mu\nu}h_{\mu\nu} - h^2\right) + 2k_{\mu}k_{\nu}g_{\alpha\beta}\left(h^{\mu\nu}h^{\alpha\beta} - h^{\mu\alpha}h^{\nu\beta}\right).
\end{aligned}
\eeq
Therefore, small and slowly changing perturbations of metric acquire a mass if the effective action contains such covariant contributions as:
\beq
    \Delta\Gamma_{\text{eff}} = \int d^Dx\sqrt{|g|}\left[\epsilon_{\text{gh}}R\frac{1}{\Box^2}R +  m_{\text{g}}^2\frac{1}{\Box}R \right].
\eeq

In general, the situation is much more intricate due to ultraviolet effects, renormalizations of cosmological constant, conformal anomalies, and other factors. For example, one of the primary contributions to induced gravity in two-dimensional space is the Mabuchi action \cite{Ferrari:2011rk, delacroixdelavalette:tel-01706737}, which originates from the integral of the Green function for the covariant Laplacian taken at coincident points.

We have not been able to solve all the puzzles that arise in this way up to this point. However, we define and analyze the quantity of effective mass as a measure of the backreaction of quantum matter immersed in the strong gravitational background for the simplest case of Bunch-Davies state in Poincaré patch of dS. As long as we treat the gravitational sector at the classical level, the notion of induced mass should not be referred to as some mass of the particle graviton but must be considered as a characteristic of matter’s behavior in the given state. For example, in the case of large positive masses, gravitational interaction is screened, and a negative squared mass corresponds to the decay of the initial external background. In particular, in the presence of classical stress-energy tensor, a negative squared mass for thermal state of matter leads to the well-known Jeans’ instability \cite{1902RSPTA.199....1J, Gorbunov:1354521, gorbunov2011introduction}. At the same time, taking into account such loop effects as secularly growing corrections  \cite{Krotov:2010ma, Polyakov:2012uc, Akhmedov:2012pa, Akhmedov:2014doa, Akhmedov:2013xka , Akhmedov:2013vka}, the stability of dS is a separate interesting question with many unresolved problems, because these contributions can drastically affect the tree-level situation for different types of quantum fields and various initial states. This is why we believe that more approaches are needed to treat this issue.

Another curious aspect of the appearance of the gauge field’s mass is the connection with the Higgs mechanism. We expect that the field of spin-$s$ swallows the Goldstone boson of spin-$(s-1)$ to acquire a mass. In Minkowski space with $\lambda \phi^4$ potential, there are two diagrams, local and non-local, which combine into a transverse structure and shift the pole of the gauge field’s propagator:
\begin{figure}[hbt!]
    \centering
    \def\svgwidth{\textwidth}
\begingroup%
  \makeatletter%
  \providecommand\color[2][]{%
    \errmessage{(Inkscape) Color is used for the text in Inkscape, but the package 'color.sty' is not loaded}%
    \renewcommand\color[2][]{}%
  }%
  \providecommand\transparent[1]{%
    \errmessage{(Inkscape) Transparency is used (non-zero) for the text in Inkscape, but the package 'transparent.sty' is not loaded}%
    \renewcommand\transparent[1]{}%
  }%
  \providecommand\rotatebox[2]{#2}%
  \newcommand*\fsize{\dimexpr\f@size pt\relax}%
  \newcommand*\lineheight[1]{\fontsize{\fsize}{#1\fsize}\selectfont}%
  \ifx\svgwidth\undefined%
    \setlength{\unitlength}{250.79328162bp}%
    \ifx\svgscale\undefined%
      \relax%
    \else%
      \setlength{\unitlength}{\unitlength * \real{\svgscale}}%
    \fi%
  \else%
    \setlength{\unitlength}{\svgwidth}%
  \fi%
  \global\let\svgwidth\undefined%
  \global\let\svgscale\undefined%
  \makeatother%
  \begin{picture}(1,0.16382336)%
    \lineheight{1}%
    \setlength\tabcolsep{0pt}%
    \put(0,0){\includegraphics[width=\unitlength,page=1]{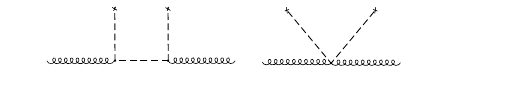}}%
    \put(0.01835877,0.04459937){\color[rgb]{0,0,0}\makebox(0,0)[lt]{\lineheight{1.25}\smash{\begin{tabular}[t]{l}$\Pi_{\mu\nu} = $\end{tabular}}}}%
    \put(0.46657316,0.04383528){\color[rgb]{0,0,0}\makebox(0,0)[lt]{\lineheight{1.25}\smash{\begin{tabular}[t]{l}$+$\end{tabular}}}}%
    \put(0.77731148,0.041074){\color[rgb]{0,0,0}\makebox(0,0)[lt]{\lineheight{1.25}\smash{\begin{tabular}[t]{l}$\propto \left(g_{\mu\nu} - \frac{k_{\mu}k_{\nu}}{k^2}\right)M_{\text{ph}}$,\end{tabular}}}}%
  \end{picture}%
\endgroup%

    \caption{\label{fig:Higgs} Higgs mechanism in Standard Model}
    \hfill
\end{figure}
\\
where crosses represent the vacuum expectation of the Higgs field, and the pole at $k^2=0$ indicates the exchange of the massless boson. If there is no potential with spontaneous symmetry breaking mechanism and our photon or graviton interacts with free field theory, we have bubble and tadpole diagrams instead of the contributions of fig.\ref{fig:Higgs}. For example, in the case of the graviton, the analogue of the Higgs mechanism will occur if a pole corresponding to the Goldstone vector appears in the non-local part of the graviton’s self-energy. This is the case when the state produced by the matter stress-energy tensor $T_{\mu\nu}(x)|0 \rangle$ has a non-zero overlap with the state of Goldstone vector, which can be easily seen if one inserts the sum over all states into the non-local part of the self-energy $\Sigma_{\mu\nu|\alpha\beta} \sim \sum\limits_{\text{state}} \langle 0|T_{\mu\nu}|\text{state}\rangle\langle\text{state}|T_{\alpha\beta}|0\rangle $. As long as the stress-energy tensor is quadratic in fields, the Higgs mechanism requires the appearance of the Goldstone vector in the tensor product of states in the matter spectrum, which is a sum of infinite-dimensional positive-weight unitary irreducible representations of the isometry group of the space under consideration \cite{Nicolai:1984hb, Fronsdal:1978vb, Porrati:2001db}. Although it is difficult to imagine that the stress-energy tensor of a free field theory can create a Goldstone vector as a bound state, this actually happens under certain conditions in AdS: e.g. the presence of the Goldstone vector in the bubble diagram is shown in \cite{Porrati:2001db, Porrati:2003sa} using the expansion of propagators at large distances.

In dS, the spectrum of states is different \cite{Penedones:2023uqc}, and we do not expect the same phenomena to occur. Moreover, the analysis in dS should be more careful as it is a non-stationary background, so one has to adopt the Schwinger-Keldysh diagrammatic technique. In particular, it was shown in \cite{Popov:2017xut} that Debye and magnetic masses of the photon in dS are zero in the maximally symmetric and analytic Bunch-Davies state. In this paper, we show that the effective mass of the tensor mode of the graviton is also zero up to a subtraction of UV divergent contact terms, which are present in Minkowsky space as well. Specifically, in Section \ref{sec:Definitions}, we describe the particular model in question and provide Schwinger-Keldysh diagrammatic technique for it. In Section \ref{sec:Effective_action}, we derive the expression for gravity’s induced action in terms of loop integrals and then use it in Section \ref{sec:Tensor_mode} to calculate the mass $m_{\text{g}}$ of the tensor mode of the graviton. We give a definition to this quantity in a manner of non-equilibrium condensed matter physics \cite{Boyanovsky:1999jh}. Finally, in Section \ref{sec:Scalar_sector}, we discuss some features that arise in the scalar sector of gravity in the effective action. First, for space-time dimension $D>2$, the loop integral for the mass of the scalar mode diverges, requiring a more careful regularization procedure that preserves the symmetries of the problem for further analysis. Second, in dS space, this mass already has a nonzero value at the classical level, with both the massive and kinetic terms entering the action with the wrong sign. This is not a problem in classical theory since scalar modes do not propagate in it. Third, we claim that divergent terms appear in a non-stationary gravitational background that are absent in the flat case. This feature should be related to a deficiency in defining effective mass as a term in the expansion of the effective action into a series and should be eliminated after resummation, so more detailed analysis needs to be carried out in subsequent studies. We separately considered the case of two-dimensional spacetime, where integrals converge and found that in dS, the effective mass of the scalar mode differs significantly from its formal value in flat space for light matter fields, indicating a significantly different response to external background in these two situations. Additionally, in Appendix \ref{App:Photon_AdS}, we show the presence of Goldstone scalar and mass of a photon in one-loop photon’s self-energy in $\text{AdS}_4$ following the spirit of work \cite{Porrati:2001db} to establish differences between field theories in AdS and dS.

\section{Preliminaries and definitions}\label{sec:Definitions}
Consider the action for gravity coupled to the real massive scalar field in $D = d+1$ dimensions:
\begin{equation}\label{eq:bare_action}
    S[g_{\mu\nu}, \phi] = -\frac{1}{16\pi G}\int d^{D}x \sqrt{|g|}\left[R + 2\Lambda\right] + \frac{1}{2}\int d^{D}x\sqrt{|g|}\left[ g^{\mu\nu}\partial_{\mu}\phi\partial_{\nu}\phi - M^2\phi^2\right],
\end{equation}
where the $\Lambda$-term is defined by the Hubble constant $H$ as $\Lambda = \frac{(D-1)(D-2)}{2}H^2$, $G$ is a Newton's constant and below we use the dimensionless mass parameter $m = \frac{M}{H}$. We will split the metric into the background in the Poincaré patch of dS and small perturbation, $h_{\mu\nu}$, over it:
\begin{equation}\label{eq:perturbed_dS}
    g_{\mu\nu} = \hat{g}_{\mu\nu} + h_{\mu\nu}, \;\; \hat{g}_{\mu\nu} = \frac{1}{H^2\eta^2}\text{diag}\left(1,-1,\ldots,-1\right),
\end{equation}
where $\eta$ is the conformal time, which is related to the inertial observer time coordinate as $\eta = \frac{1}{H}e^{-Ht}$. Below we will also use the perturbation with raised indices $h^{\mu\nu} \overset{\text{def}}{=} \hat{g}^{\mu\alpha}\hat{g}^{\nu\beta}h_{\alpha\beta}$ and the rescaled field $\mathscr{h}_{\mu\nu} = H^2\eta^2 h_{\mu\nu}$, such that $g_{\mu\nu} = \frac{1}{H^2\eta^2}\left[\gamma_{\mu\nu} + \mathscr{h}_{\mu\nu}\right],\;\gamma_{\mu\nu} = \text{diag}\left(1,-1,\ldots,-1\right)$. The field $\mathscr{h}_{\mu\nu}$ is a more appropriate variable for the problem in question, e.g. the equations of motion for the naive linearized massive gravity take the form of the usual Klein-Gordon equation for the fields, obtained from the components of $\mathscr{h}_{\mu\nu}$ by means of linear operations \cite{Jaccard:2012ut}. We consider gravity as classical and quantize only the scalar field.
\subsection{The quantization of the scalar field}
We quantize the scalar field in the standard way using the creation and annihilation operators with the canonical commutation relations:
\beq \label{eq:field_decomposition}
    \begin{aligned}
        \phi(\eta,\mathbf{x})=\int\dfrac{d^{D-1}\mathbf{p}}{(2\pi)^{D-1}}\bigg[
\widehat{a}_{\mathbf{p}}f_{\mathbf{p}}(\eta)e^{i\mathbf{px}}+
\widehat{a}^{\dagger}_{\mathbf{p}}f^{*}_{\mathbf{p}}(\eta)e^{-i\mathbf{px}}
\bigg], \quad 
\left[\widehat{a}_{\mathbf{p}},\widehat{a}_{\mathbf{q}}^\dagger 
\right] = (2\pi)^{D-1}\delta(\mathbf{p}-\mathbf{q}),
\\
f_{\mathbf{p}}(\eta) = H^{\frac{D-2}{2}}\eta^{\frac{D-1}{2}}h_{\nu}\left(p\eta\right),\;p\equiv |\mathbf{p}|.
    \end{aligned}
\eeq
Here $h_{\nu}(p\eta)$ can be expressed in terms of the Hankel function of the first kind $H_{\nu}^{(1)}(z)$ for complementary $\left(m<\frac{D-1}{2}\right)$ and principle $\left(m>\frac{D-1}{2}\right)$ series as follows:
\beq \label{eq:harmonics}
\begin{aligned}
        h_{\nu}\left(p\eta\right) &= \frac{\sqrt{\pi}}{2}e^{-\frac{\pi}{2}\nu}H_{i\nu}^{(1)}\left(p\eta\right), \;\nu = \sqrt{m^2-\frac{(D-1)^2}{4}} \;\;\text{(principal series)}, \\
    h_{\nu}\left(p\eta\right) &= \frac{\sqrt{\pi}}{2}H_{\nu}^{(1)}\left(p\eta\right),\; \nu = \sqrt{\frac{(D-1)^2}{4}-m^2} \;\;\text{(complementary series)},
\end{aligned}
\eeq
so that the mode functions $f_{\mathbf{p}}(\eta)$ obey the classical equation of motion
\begin{equation}\label{eq:EOM_for_harmonics}
    \nabla_{\eta}\partial_{\eta}f_{\mathbf{p}}(\eta) + \left(p^2+\frac{m^2}{\eta^2}\right)f_{\mathbf{p}}(\eta) = 0, \quad \nabla_{\eta} \equiv \partial_{\eta} - \frac{D-2}{\eta}.
\end{equation}
Note, that by choosing the harmonics in the form (\ref{eq:harmonics}) and by the condition $\widehat{a}_{\mathbf{p}}\left|\text{BD}\right\rangle = 0$ we specify the Bunch-Davies state of the scalar field theory in the Poincaré patch of $dS_D$ -- we will stick to this initial state throughout this paper as it preserves the highest number of symmetries in loop calculations \cite{Akhmedov:2013vka}, while the effects of various nontrivial initial states will be considered elsewhere. Next, in order to construct the Schwinger-Keldysh diagrammatic technique, it is appropriate to introduce the fields after the Keldysh rotation:
\begin{equation}\label{eq:Keldysh_rotation}
    \phi_{cl} = \frac{\phi_{+} + \phi_{-}}{2}, \; \phi_{q} = \phi_{+} - \phi_{-}; \;\;
    h_{cl}^{\mu\nu} = \frac{h_{+}^{\mu\nu} + h_{-}^{\mu\nu}}{2}, \; h_{q}^{\mu\nu} = h_{+}^{\mu\nu} - h_{-}^{\mu\nu}.
\end{equation}
Here ``$+$''- and ``$-$''-parts are attributed to the upper and lower branches of the Keldysh contour $\mathcal{C}$ on $t$--plane:
\\
\begin{figure}[hbt!]
    \centering
    \def\svgwidth{\textwidth}
\begingroup%
  \makeatletter%
  \providecommand\color[2][]{%
    \errmessage{(Inkscape) Color is used for the text in Inkscape, but the package 'color.sty' is not loaded}%
    \renewcommand\color[2][]{}%
  }%
  \providecommand\transparent[1]{%
    \errmessage{(Inkscape) Transparency is used (non-zero) for the text in Inkscape, but the package 'transparent.sty' is not loaded}%
    \renewcommand\transparent[1]{}%
  }%
  \providecommand\rotatebox[2]{#2}%
  \newcommand*\fsize{\dimexpr\f@size pt\relax}%
  \newcommand*\lineheight[1]{\fontsize{\fsize}{#1\fsize}\selectfont}%
  \ifx\svgwidth\undefined%
    \setlength{\unitlength}{437.70141121bp}%
    \ifx\svgscale\undefined%
      \relax%
    \else%
      \setlength{\unitlength}{\unitlength * \real{\svgscale}}%
    \fi%
  \else%
    \setlength{\unitlength}{\svgwidth}%
  \fi%
  \global\let\svgwidth\undefined%
  \global\let\svgscale\undefined%
  \makeatother%
  \begin{picture}(1,0.24825987)%
    \lineheight{1}%
    \setlength\tabcolsep{0pt}%
    \put(0,0){\includegraphics[width=\unitlength,page=1]{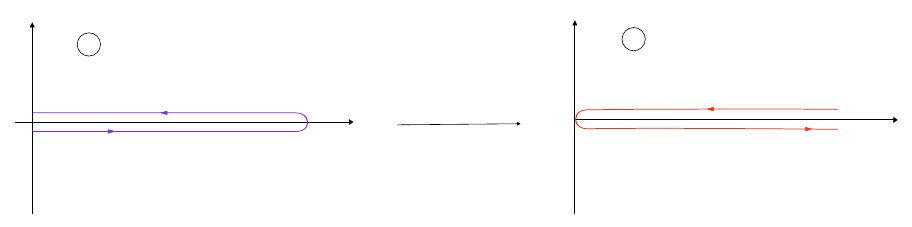}}%
    \put(0.09314723,0.19209487){\color[rgb]{0,0,0}\makebox(0,0)[lt]{\lineheight{1.25}\smash{\begin{tabular}[t]{l}$t$\end{tabular}}}}%
    \put(0.68909522,0.20069609){\color[rgb]{0,0,0}\makebox(0,0)[lt]{\lineheight{1.25}\smash{\begin{tabular}[t]{l}$\eta$\end{tabular}}}}%
    \put(0.34106729,0.09976802){\color[rgb]{0,0,0}\makebox(0,0)[lt]{\lineheight{1.25}\smash{\begin{tabular}[t]{l}$\infty$\end{tabular}}}}%
    \put(0.61564673,0.11059838){\color[rgb]{0,0,0}\makebox(0,0)[lt]{\lineheight{1.25}\smash{\begin{tabular}[t]{l}$0$\end{tabular}}}}%
    \put(0.20765663,0.13457078){\color[rgb]{0,0,0}\makebox(0,0)[lt]{\lineheight{1.25}\smash{\begin{tabular}[t]{l}$\mathcal{C}$\end{tabular}}}}%
    \put(0.45340578,0.07680642){\color[rgb]{0,0,0}\makebox(0,0)[lt]{\lineheight{1.25}\smash{\begin{tabular}[t]{l}$\eta = \frac{1}{H}e^{-Ht}$\end{tabular}}}}%
  \end{picture}%
\endgroup%

    \caption{\label{fig:Contour} Keldysh contour on $t-$ and $\eta-$plane}
    \hfill
\end{figure}
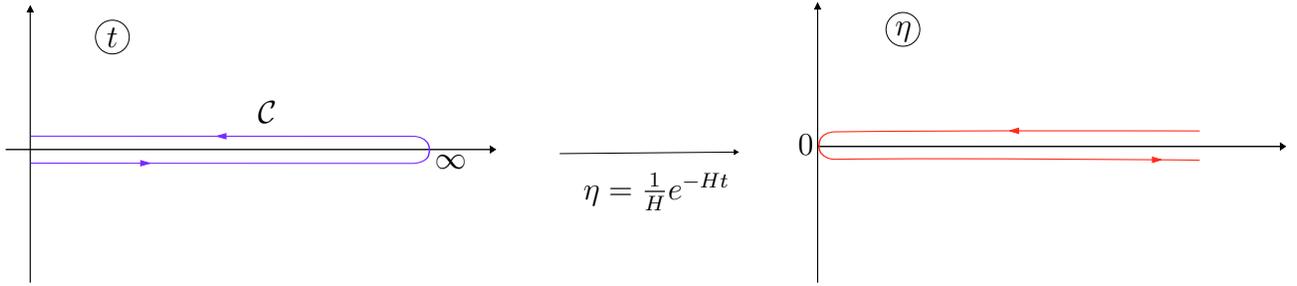
\\
The corresponding propagators of the scalar field in these notations have the form ($\mathcal{T}_{\mathcal{C}}$ is the ordering operator along the contour $\mathcal{C}$ on the fig.\ref{fig:Contour}):
\\
\begin{equation}\label{eq:dS_propagators}
\begin{aligned}
    G(x,x') &=  \langle \mathcal{T}_{\mathcal{C}}\varphi(x)\varphi(x')\rangle = F(x,x') - \frac{i}{2}\text{sign}_{\mathcal{C}}(\eta - \eta')\rho(x,x'),
    \\
    \langle \phi_{cl}(x)\phi_{cl}(y)\rangle &= F(x,y), \;\; \langle \phi_{q}(x)\phi_{cl}(y)\rangle = i\theta(\eta - \eta')\rho(x,y),
    \\
    \langle \phi_{cl}(x)\phi_{q}(y)\rangle &= -i\theta(\eta' - \eta)\rho(x,y), \;\; \langle \phi_{q}(x)\phi_{q}(y)\rangle = 0,
\end{aligned}
\end{equation}
\\
where the sign function $\text{sign}_{\mathcal{C}}$ is implemented along the contour $\mathcal{C}$, $F(x,y) = \frac{1}{2}\left\langle\{\phi(x), \phi(y)\}\right\rangle$ and $\rho(x,y) = i\left\langle[\phi(x),\phi(y)]\right\rangle$ are the Keldysh and spectral functions respectively \cite{Serreau:2013psa}. In the following discussion we will use the spatially Fouriér-transformed propagators:
\begin{equation}\label{eq:Fourier_propagators}
\begin{aligned}
    F(\mathbf{k}|\eta,\eta') &= \int d^{d}\mathbf{x}F(\eta,\eta',|\mathbf{x}-\mathbf{y}|)e^{-i\mathbf{k}(\mathbf{x}-\mathbf{y})} = \text{Re}\left\{f_{\mathbf{p}}(\eta)f_{\mathbf{p}}^{*}(\eta')\right\},
    \\
    \rho(\mathbf{k}|\eta,\eta') &= \int d^{d}\mathbf{x}\rho(\eta,\eta',|\mathbf{x}-\mathbf{y}|)e^{-i\mathbf{k}(\mathbf{x}-\mathbf{y})} = -2\text{Im}\left\{f_{\mathbf{p}}(\eta)f_{\mathbf{p}}^{*}(\eta')\right\},
\end{aligned}
\end{equation}
because the state that we consider is spatially homogeneous. Also it is worth noting here that the commutation relation $[\phi(x),\pi(y)] = i\delta^{(d)}(\mathbf{x}-\mathbf{y})$ with the canonical momentum $\pi(\mathbf{x},t) = \sqrt{|g|}g^{00}(t)\partial_{t}\phi(\mathbf{x},t)$ implies the following property of the spectral function:
\begin{equation}\label{eq:canonical_commutator}
    \partial_{\eta}\rho(\mathbf{k}|\eta,\eta')\bigg|_{\eta=\eta'} = -H^{D-2}\eta^{D-2}, \quad 
    \partial_{\eta'}\rho(\mathbf{k}|\eta,\eta')\bigg|_{\eta=\eta'} = H^{D-2}\eta^{D-2},
\end{equation}
while the causality requires $\rho(\mathbf{k}|\eta,\eta) = \partial_{\eta}\partial_{\eta'}\rho(\mathbf{k}|\eta,\eta')\bigg|_{\eta=\eta'} = 0$.
\subsection{Effective equation of motion}
The propagators (\ref{eq:dS_propagators}) allow us to find perturbatively the Keldysh effective action $\Gamma_{\text{eff}}[h_{cl},h_{q}]$, which is a powerful tool to study dynamics of non-equilibrium systems \cite{Berges:2004yj, Wang:2022mvv, Ai:2021gtg, kamenev_2011}. To accomplish this, we extend the integration in (\ref{eq:bare_action}) onto the contour $\mathcal{C}$, change the fields according to (\ref{eq:Keldysh_rotation}), expand the functional integral over the matter fields in powers of $h_{\mu\nu}$ and calculate loop integrals using the propagators (\ref{eq:dS_propagators}). The contributions we are interested in are as follows:
\\
\begin{figure}[hbt!]
    \centering
    \def\svgwidth{\textwidth}
\begingroup%
  \makeatletter%
  \providecommand\color[2][]{%
    \errmessage{(Inkscape) Color is used for the text in Inkscape, but the package 'color.sty' is not loaded}%
    \renewcommand\color[2][]{}%
  }%
  \providecommand\transparent[1]{%
    \errmessage{(Inkscape) Transparency is used (non-zero) for the text in Inkscape, but the package 'transparent.sty' is not loaded}%
    \renewcommand\transparent[1]{}%
  }%
  \providecommand\rotatebox[2]{#2}%
  \newcommand*\fsize{\dimexpr\f@size pt\relax}%
  \newcommand*\lineheight[1]{\fontsize{\fsize}{#1\fsize}\selectfont}%
  \ifx\svgwidth\undefined%
    \setlength{\unitlength}{536.20963029bp}%
    \ifx\svgscale\undefined%
      \relax%
    \else%
      \setlength{\unitlength}{\unitlength * \real{\svgscale}}%
    \fi%
  \else%
    \setlength{\unitlength}{\svgwidth}%
  \fi%
  \global\let\svgwidth\undefined%
  \global\let\svgscale\undefined%
  \makeatother%
  \begin{picture}(1,0.11837124)%
    \lineheight{1}%
    \setlength\tabcolsep{0pt}%
    \put(0,0){\includegraphics[width=\unitlength,page=1]{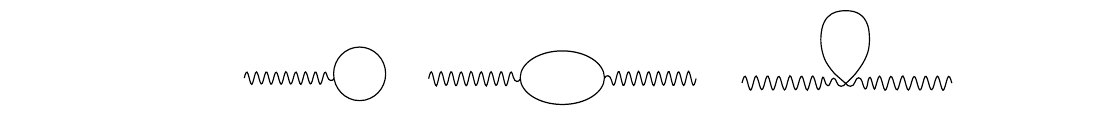}}%
    \put(0.07127421,0.04492134){\color[rgb]{0,0,0}\makebox(0,0)[lt]{\lineheight{1.25}\smash{\begin{tabular}[t]{l}$\Gamma_{\text{eff}}$\end{tabular}}}}%
    \put(0.11539138,0.04444519){\color[rgb]{0,0,0}\makebox(0,0)[lt]{\lineheight{1.25}\smash{\begin{tabular}[t]{l}$=$\end{tabular}}}}%
    \put(0.14921132,0.04464401){\color[rgb]{0,0,0}\makebox(0,0)[lt]{\lineheight{1.25}\smash{\begin{tabular}[t]{l}$S_{\text{cl}}$\end{tabular}}}}%
    \put(0.35454602,0.04554303){\color[rgb]{0,0,0}\makebox(0,0)[lt]{\lineheight{1.25}\smash{\begin{tabular}[t]{l}$+$\end{tabular}}}}%
    \put(0.63367228,0.04191763){\color[rgb]{0,0,0}\makebox(0,0)[lt]{\lineheight{1.25}\smash{\begin{tabular}[t]{l}$+$\end{tabular}}}}%
    \put(0.18595753,0.04542815){\color[rgb]{0,0,0}\makebox(0,0)[lt]{\lineheight{1.25}\smash{\begin{tabular}[t]{l}$+$\end{tabular}}}}%
    \put(0.85999879,0.0378597){\color[rgb]{0.04313725,0.04313725,0.04313725}\makebox(0,0)[lt]{\lineheight{1.25}\smash{\begin{tabular}[t]{l}$.$\end{tabular}}}}%
  \end{picture}%
\endgroup%

    \caption{\label{fig:Effective_action} Effective action}
    \hfill
\end{figure}
\\
As we will see, all the diagrams on the fig.\ref{fig:Effective_action} are important for the effective action to be gauge invariant in the order under consideration. Also there can be some additional counterterms $\delta_{\text{ren}}\Gamma$ needed to cure the UV divergences in these loops -- we will discuss them in the next section and show, that the first diagram on the fig.\ref{fig:Effective_action} can be subtracted by the term $\delta_{\Lambda}\Gamma_{\text{eff}}$, which renormalizes the cosmological constant. 

The graviton equation of motion (EOM) follows from the effective action $\Gamma_{\text{eff}}\left[h_{cl}, h_{q}\right]$ as
\begin{equation}\label{eq:EOM_from_Gamma}
    \frac{\delta}{\delta h_{q}} \Gamma_{\text{eff}}\left[h_{cl}, h_{q}\right]\bigg|_{h_{q} = 0} = 0.
\end{equation}
To derive these equations we need the following interacting parts in the action (\ref{eq:bare_action}):
\begin{equation}\label{interacting_action}
\Delta S = -\int d^{D}x\sqrt{|\hat{g}|} h_{cl}^{\mu\nu}T_{\mu\nu}^{cl-q} - \frac{1}{2}\int d^{D}x\sqrt{|\hat{g}|} h_{q}^{\mu\nu}T_{\mu\nu}^{cl-cl} + \int d^{D}x\sqrt{|\hat{g}|} h_{cl}^{\mu\nu}\Gamma_{\mu\nu|\alpha\beta}h_{q}^{\alpha\beta},
\end{equation}
where
\begin{equation}\label{eq:Stress-energy_tensor}
    T_{\mu\nu}^{cl-q} = -\frac{1}{2}\hat{g}_{\mu\nu}\left(\hat{g}^{\lambda\omega}\partial_{\lambda}\phi^{cl}\partial_{\omega}\phi^{q} - m^2\phi^{cl}\phi^{q} \right) + \partial_{\mu}\phi^{cl}\partial_{\nu}\phi^{q},
\end{equation}
\begin{equation}\label{eq:bare_tadpole}
    \Gamma_{\mu\nu|\alpha\beta} = \frac{1}{8}\left( \hat{g}_{\alpha\beta}\hat{g}_{\mu\nu} - 2\hat{g}_{\mu\alpha}\hat{g}_{\nu\beta}\right)\left[\hat{g}^{\lambda\omega}\partial_{\lambda}\phi^{cl}\partial_{\omega}\phi^{cl} - m^2\phi^{cl\;2}  \right] - \frac{1}{2}\hat{g}_{\alpha\beta}\partial_{\mu}\phi^{cl}\partial_{\nu}\phi^{cl} + \frac{1}{2}\hat{g}_{\mu\alpha}\partial_{\beta}\phi^{cl}\partial_{\nu}\phi^{cl}
\end{equation}
and the same for $T_{\mu\nu}^{cl-cl}$ replacing $q\rightarrow cl$ in (\ref{eq:Stress-energy_tensor}). Note that the bare correlation funcions (\ref{eq:dS_propagators}) contain theta-functions, while the bubble diagram on the fig.\ref{fig:Effective_action} has derivatives over time in the vertices as they appear in the stress-energy tensor (\ref{eq:Stress-energy_tensor}). Hence, there can be delta-functions in the bubble diagram\footnote{Note, that in the operator formalism time derivatives do not commute with the time-ordering operator: $\left\langle\mathcal{T}\partial_t \hat{A}(t)\hat{B}(t')\right\rangle \neq \partial_t\left\langle\mathcal{T}\hat{A}(t)\hat{B}(t')\right\rangle$. However, if time derivatives appear in vertices, there additional non-covariant terms emerge in the interaction Hamiltonian, which restore the accordance with the functional-integral approach, where one can carry the time derivatives through the functional integral \cite{itzykson2012quantum}.}, so we will collect these local contributions $\Delta\Pi^{\text{loc}}_{\mu\nu|\alpha\beta}$ to the total polarization operator along with the tadpole diagram into the one expression $\Pi_{\mu\nu|\alpha\beta}^{\text{loc}}$ in what follows, while we will denote by $\Pi_{\mu\nu|\alpha\beta}^{\text{bub}}$ the non-local contributions, where all the derivatives in vertices act only on the Keldysh and spectral functions of the propagators (\ref{eq:dS_propagators}) in this diagram. Then the effective EOM in the momentum space over the $d$ space-coordinates has the form:
\begin{gather}
    \frac{1}{16\pi G}\widehat{EOM}_{\mu\nu|\alpha\beta}h^{\mu\nu}(\mathbf{k},\eta) - \nonumber
    \\
    - \frac{1}{2}\int_{\eta}^{\infty}\frac{d\eta'}{H^D \eta'^{D}} \Pi_{\mu\nu|\alpha\beta}^{\text{bub}}\left(\mathbf{k}|\eta', \eta\right) h^{\mu\nu}(\mathbf{k},\eta') + \Pi_{\mu\nu|\alpha\beta}^{\text{loc}}\left(\eta\right) h^{\mu\nu}(\mathbf{k},\eta) = \frac{1}{2}\left\langle T_{\alpha\beta}^{cl-cl}\right\rangle,
\label{eq:Eff_EOM}
\end{gather}
where the ``source''-term on the RHS corresponds to the first tadpole diagram on the fig.\ref{fig:Effective_action} and
\begin{equation}\label{eq:Local_contribution}
    \Pi_{\mu\nu|\alpha\beta}^{\text{loc}}\left(\eta\right) = -\frac{1}{2}\Delta \Pi_{\mu\nu|\alpha\beta}^{\text{loc}}\left(\eta\right) + \Pi_{\mu\nu|\alpha\beta}^{\text{tad}}\left(\eta\right),
\end{equation}
\begin{equation}\label{eq:Bubble_contribution}
    \Pi_{\mu\nu|\alpha\beta}^{\text{bub}} + \Delta \Pi_{\mu\nu|\alpha\beta}^{\text{loc}} = \left\langle \frac{T_{\mu\nu}^{cl-q} + T_{\nu\mu}^{cl-q}}{2} T_{\mu\nu}^{cl-cl}\right\rangle.
\end{equation} 
The operator $\widehat{EOM}_{\mu\nu|\alpha\beta}$ in the equation (\ref{eq:Eff_EOM}) appears due to the Einstein-Hilbert part of the action (\ref{eq:bare_action}). Namely, following \cite{Gorbunov:1354521, gorbunov2011introduction}, we split the metric perturbation onto the spiral components:
\beq\label{eq:spiral_decomposition}
\begin{aligned}
    \mathscr{h}_{00} &= 2\Phi, 
    \\
    \mathscr{h}_{0k} &= ik_{k}Z + Z_{k}^{T},
    \\
    \mathscr{h}_{kl} &= -2\Psi\delta_{kl} - 2k_{k}k_{l}E + i(k_{k}W_{j}^{T} + k_{l}W_{i}^{T}) +  \mathscr{h}_{kl}^{TT},
\end{aligned}
\eeq
where $k_{k}Z_{k}^{T} = k_{k}W_{k}^{T} = k_{k}\mathscr{h}_{kl}^{TT} = 0$ and $\mathscr{h}_{kk}^{TT} = 0$. We will work in the gauge $\mathscr{h}_{0k} = 0$. In this gauge the linearized Einsein's tensor $G_{\mu\nu}^{(1)} = R_{\mu\nu}^{(1)} -\frac{1}{2}\hat{g}_{\mu\nu}R^{(1)} + (D-1)H^2 h_{\mu\nu}$ in arbitrary dimension has the form:
\beq \label{eq:Linearized_Einstein_EOM}
\begin{aligned}
    G_{00}^{(1)} &= -\frac{(D-1)(D-2)}{2\eta^2}\mathscr{h}_{00} + \frac{D-2}{2\eta}\partial_{\eta}\mathscr{h}_{kk} + \frac{1}{2}\partial^2_{l}\mathscr{h}_{kk} - \frac{1}{2}\partial_{l}\partial_{k}\mathscr{h}_{kl},
    \\
    G_{0i}^{(1)} &= -\frac{D-2}{2\eta}\partial_{i}\mathscr{h}_{00} - \frac{1}{2}\partial_{\eta}\partial_{k}\mathscr{h}_{ki} + \frac{1}{2}\partial_{\eta}\partial_{i}\mathscr{h}_{kk},
    \\
   G_{ij}^{(1)} &= -\frac{1}{2}\partial_{i}\partial_{j}\mathscr{h}_{00} + \frac{1}{2}\partial_{i}\partial_{j}\mathscr{h}_{kk} - \frac{1}{2}\partial_{\eta}^{2}\mathscr{h}_{ij} + \frac{D-2}{2\eta}\partial_{\eta}\mathscr{h}_{ij} + \frac{1}{2}\partial^2_{k}\mathscr{h}_{ij} -
    \\
    &- \frac{1}{2}\left(  \partial_{i}\partial_{k}\mathscr{h}_{kj} + \partial_{j}\partial_{k}\mathscr{h}_{ki}\right)+\delta_{ij}\bigg[\frac{(D-1)(D-2)}{2\eta^2}\mathscr{h}_{00} - \frac{D-2}{2\eta}\partial_{\eta}\mathscr{h}_{00} + \frac{1}{2}\partial^2_{k}\mathscr{h}_{00} +
    \\
    &\qquad\qquad\qquad\qquad\qquad\qquad\qquad+ \frac{1}{2}\partial^2_{\eta}\mathscr{h}_{kk} - \frac{D-2}{2\eta}\partial_{\eta}\mathscr{h}_{kk} - \frac{1}{2}\partial_{l}^2 \mathscr{h}_{kk} + \frac{1}{2}\partial_{l}\partial_{k}\mathscr{h}_{kl} \bigg],
\end{aligned}
\eeq
which defines the action of the operator $\widehat{EOM}_{\mu\nu|\alpha\beta}$ in the first line of (\ref{eq:Eff_EOM}). We will use the equations (\ref{eq:Eff_EOM})--(\ref{eq:Linearized_Einstein_EOM}) to properly define the notion of the induced mass in the following sections. 
\subsection{Implications of de Sitter isometries}
In Bunch-Davies state, after the subtraction of the $\Lambda$-renormalization counterterm $\delta_{\Lambda}\Gamma_{\text{eff}}$ from $\Gamma_{\text{eff}}$, we are left with the equation of motion of the form (\ref{eq:Eff_EOM}), but with no ``source''-term on the RHS and with renormalized local part of the polarizatrion operator. This equation is invariant under the gauge transformation in the zeroth order in the metric perturbations $\delta_{\xi}h^{\mu\nu} = -\hat{g}^{\mu\lambda}\hat{\nabla}_{\lambda}\xi^{\nu} - \hat{g}^{\nu\lambda}\hat{\nabla}_{\lambda}\xi^{\mu}$. In order to obtain a general form of the linear equation $\widehat{\mathscr{D}}_{\mu\nu|\alpha\beta}h^{\alpha\beta}=0$, which respects both the dS isometry group $SO(1,D)$ and gauge invariance, we will use the Lichnerowicz operator $\Delta_{L}$. It acts on the tensor, vector and scalar fields in the following way:
\begin{equation}\label{eq:lichneriwicz_action}
    \begin{aligned}
        \Delta_{L}h_{\mu\nu} &= -\hat{\Box}h_{\mu\nu} - 2\hat{R}_{\mu\alpha\nu\beta}h^{\alpha\beta} + \hat{R}^{\alpha}_{\mu}h_{\nu\alpha} + \hat{R}^{\alpha}_{\nu}h_{\mu\alpha},
        \\
        \Delta_{L}V_{\mu} &= \left(-\hat{\Box} - \frac{2\Lambda}{D-2}\right)V_{\mu},
        \\
        \Delta_{L}\varphi &= -\hat{\Box}\varphi,
    \end{aligned}
\end{equation}
where $\hat{R}_{\mu\alpha\nu\beta}, \; \hat{R}_{\mu\nu}$ and $\hat{\Box}$ are Riemann tensor, Ricci tensor and the covariant Laplacian on the dS background correspondingly. The action of $\Delta_{L}$ in dS commutes with the covariant derivatives, as explained e.g. in \cite{Lichnerowicz2018RepublicationOP, GIBBONS197890, Porrati:2001db}. Then we can seek for the operator $\widehat{\mathscr{D}}_{\mu\nu|\alpha\beta}$ in the explicitly dS invariant form:
\beq\label{eq:general_inv_operator}
\begin{aligned}
    \widehat{\mathscr{D}}_{\mu\nu|\alpha\beta}h^{\alpha\beta} = A(\Delta_L)h_{\mu\nu} + \frac{1}{2}B(\Delta_L)\left[ \hat{\nabla}_{\mu}\hat{\nabla}^{\lambda}h_{\nu\lambda} + \hat{\nabla}_{\nu}\hat{\nabla}^{\lambda}h_{\mu\lambda}\right] + 
    \\
    + C(\Delta_L)\hat{\nabla}_{\mu}\hat{\nabla}_{\nu}\hat{\nabla}^{\alpha}\hat{\nabla}^{\beta}h_{\alpha\beta} + D(\Delta_L)\hat{\nabla}_{\mu}\hat{\nabla}_{\nu}h^{\alpha}_{\alpha} + E(\Delta_L)\hat{g}_{\mu\nu}h^{\alpha}_{\alpha} + F(\Delta_L)\hat{g}_{\mu\nu}\hat{\nabla}^{\alpha}\hat{\nabla}^{\beta}h_{\alpha\beta},
\end{aligned}
\eeq
where $A,\;B,\;C,\;D,\;E,\;F$ are integro-differential operators, which can be expressed in terms of $\Delta_L$ and its Green functions. Also we set $D = F$ immediately due to the required symmetry under the switching of the pairs of indices $(\mu\nu)\leftrightarrow(\alpha\beta)$. 

Below we show that the invariant one-loop corrected effective equation of motion can include only two independent operators, which we denote as $\widehat{\mathscr{P}}^{tt}_{\mu\nu|\alpha\beta},\;\widehat{\mathscr{P}}^{s}_{\mu\nu|\alpha\beta}$. They are associated with the projectors onto the transverse traceless part of the graviton and onto the scalar mode of the graviton correspondingly. Namely, one can verify, using the explicit expressions given below in (\ref{eq:projector_on_htt}) and (\ref{eq:projector_on_scalar}), that the operator $\widehat{EOM}_{\mu\nu|\alpha\beta}$ from (\ref{eq:Eff_EOM}) can be written as
\begin{equation}\label{eq:cl_EOM_dS_inv}
\begin{aligned}
     \widehat{EOM}_{\mu\nu|\alpha\beta} &= 
     \\
     = \bigg(\Delta_L &+ 2\left(D-1\right)H^2\bigg)\left[\widehat{\mathscr{P}}^{tt}_{\mu\nu|\alpha\beta} - \frac{D-2}{D-1}\frac{\left( \Delta_L + \left(D-1\right)H^2 \right)^2}{\left(\Delta_L + DH^2\right)\left(\Delta_L + 2\left(D-1\right)H^2 \right)}\widehat{\mathscr{P}}^{s}_{\mu\nu|\alpha\beta} \right],
\end{aligned}
\end{equation}
so that the effective linear EOM of the form $\widehat{\mathscr{D}}_{\mu\nu|\alpha\beta}h^{\alpha\beta}=0$ is as follows:
\begin{equation}\label{eq:Eff_EOM_dS_inv}
    \widehat{EOM}_{\mu\nu|\alpha\beta}h^{\alpha\beta} + A\left(\Delta_L\right)\widehat{\mathscr{P}}^{tt}_{\mu\nu|\alpha\beta}h^{\alpha\beta} + E\left(\Delta_L\right)\widehat{\mathscr{P}}^{s}_{\mu\nu|\alpha\beta}h^{\alpha\beta} = 0.
\end{equation}

Indeed, although we have 5 independent coefficients in (\ref{eq:general_inv_operator}), the requirement of gauge invariance implies three more constraints:
\begin{equation}\label{eq:gauge_inv_constraints}
    \begin{cases}
        2A - \left(\Delta_L + \frac{4\Lambda}{D-2}\right)B = 0,
        \\
        B + 2D - 2C\left(\Delta_L + \frac{2\Lambda}{D-2}\right) = 0,
        \\
        E - D\left(\Delta_L + \frac{2\Lambda}{D-2}\right) = 0,
    \end{cases}
\end{equation}
so we are left with 2 independent coefficients and, therefore, two independent dS invariant and gauge invariant tensor structures, which act on the $h_{\mu\nu}$. The first structure for the projection onto the transverse traceless part of the graviton $h^{tt}_{\mu\nu} = \widehat{\mathscr{P}}^{tt}_{\mu\nu|\alpha\beta}h^{\alpha\beta}$ can be fixed by the two additional conditions $A = 1$ and $\hat{g}^{\mu\nu}h^{tt}_{\mu\nu} = \hat{g}^{\mu\nu}\widehat{\mathscr{P}}^{tt}_{\mu\nu|\alpha\beta}h^{\alpha\beta} = 0$. These additional constraints lead to the following set of the coefficients for this  projector (we express the cosmological constant through the Hubble parameter):
\begin{equation}\label{eq:projector_on_htt}
    \begin{aligned}
        B = \frac{2}{\Delta_L + 2\left(D-1\right)H^2}, \quad C = \frac{\frac{D-2}{D-1}}{\left(\Delta_L + DH^2\right)\left(\Delta_L + 2\left(D-1\right)H^2\right)},
        \\
        F = -\frac{1}{D-1}\frac{1}{\Delta_L + DH^2},\quad E = -\frac{1}{D-1}\frac{\Delta_L +\left(D-1\right)H^2}{\Delta_L + DH^2}.
    \end{aligned}
\end{equation}
The second independent operator $\widehat{\mathscr{P}}^{s}_{\mu\nu|\alpha\beta}$ can be written in the following simple form:
\begin{equation}\label{eq:projector_on_scalar}
    \widehat{\mathscr{P}}^{s}_{\mu\nu|\alpha\beta} = \left(\hat{g}_{\mu\nu} - \frac{\hat{\nabla}_{\mu}\hat{\nabla}_{\nu}}{-\Delta_L-\left(D-1\right)H^2}\right)\left(\hat{g}_{\alpha\beta} - \frac{\hat{\nabla}_{\alpha}\hat{\nabla}_{\beta}}{-\Delta_L-\left(D-1\right)H^2}\right).
\end{equation}
A few comments must be given about the equation (\ref{eq:Eff_EOM_dS_inv}). First, we stress that $A\left(\Delta_L\right)$ and $E\left(\Delta_L\right)$ are actually integro-differential operators. Second, with the use of an intuition of flat space where $\Delta_L \sim k^2$ we can observe that the IR behaviour of $A\left(\Delta_L\right)$ and $E\left(\Delta_L\right)$ provides us with the coefficients $m_{\text{g}}$ and $\epsilon_{\text{gh}}$ in (\ref{eq:massive_gravity}). Hence, when all the symmetries are respected during our operations, it suffices to calculate the effective mass, e.g., only for the sector of gravitational perturbations $\mathscr{h}^{TT}_{\mu\nu}$ to restore the whole ``Fierz-Pauli term'' in the induced gravity at large distances. Finally, the statements of this subsection are strictly working well exclusively for Bunch-Davies initial state of the matter and for Poincaré patch of dS, because in global dS the isometries are broken at the loop-level \cite{Krotov:2010ma, Akhmedov:2012dn}.
\section{Effective action}\label{sec:Effective_action}
In this section we find the expression for the effective action $\Gamma_{\text{eff}}$. The analysis of the non-local part $\Pi_{\mu\nu|\alpha\beta}^{\text{bub}}$ is given in the Appendix \ref{subSec:Bubble}.

In order to obtain the expression for the tadpole diagram we average the second-order term (\ref{eq:bare_tadpole}) over the rotationally invariant state and get
\beq\label{eq:tadpole}
\begin{aligned}
    \Pi_{00|00}^{\text{tad}}
    &= \frac{1}{H^2\eta^2}\frac{1}{2}\int_{p}\partial_{\eta}\partial_{\eta'}F\bigg|_{\eta' = \eta} - \frac{1}{8H^4\eta^4}\left\langle \mathscr{L}(\eta)\right\rangle =: \frac{1}{H^2\eta^2}\pi_{1}(\eta),
    \\
    \Pi_{00|0k}^{\text{tad}} &= \Pi_{0i|kl}^{\text{tad}} = 0,
    \\
    \Pi_{00|kl}^{\text{tad}} &= -\frac{1}{8H^4\eta^4}\left\langle \mathscr{L}(\eta)\right\rangle\delta_{kl} +\frac{1}{4H^2\eta^2}\int_{p}\left[\partial_{\eta}\partial_{\eta'}F\bigg|_{\eta' = \eta} - \frac{1}{D-1}p^2F\right]\delta_{kl} =: \frac{1}{H^2\eta^2}\pi_{2}(\eta)\delta_{kl},
    \\
    \Pi_{0i|0l}^{\text{tad}} &= -\frac{1}{H^2\eta^2}\pi_{2}(\eta)\delta_{il},
    \\
    \Pi_{ij|kl}^{\text{tad}} &= \frac{1}{8}\left[\frac{1}{H^4\eta^4} \left\langle \mathscr{L}(\eta)\right\rangle + \frac{1}{H^2\eta^2}\frac{4}{D-1}\int_{p}p^2 F  \right]\times\left( \delta_{ij}\delta_{kl} - \delta_{ik}\delta_{jl} - \delta_{il}\delta_{jk} \right) =
    \\
    &\qquad\qquad\qquad\qquad\qquad\qquad\qquad\qquad\qquad\quad=: \frac{1}{H^2\eta^2}\pi_{3}(\eta)\left( \delta_{ij}\delta_{kl} - \delta_{ik}\delta_{jl} - \delta_{il}\delta_{jk} \right),
\end{aligned}
\eeq
where $\int_{p} = \int\frac{d^{D-1}\mathbf{p}}{(2\pi)^{D-1}}$, the Keldysh function $F(\mathbf{p}|\eta,\eta')$ (for brevity we drop the arguments in the expressions above under the integrals) is taken at coincident points $\eta=\eta'$ and we have introduced the averaged Lagrangian
\begin{equation}\label{eq:lagrangian_averaged}
\langle \mathscr{L}(\eta)\rangle = H^2\eta^2\int_{p}\left[ \partial_{\eta}\partial_{\eta'}F\bigg|_{\eta' = \eta} - \left(p^2 +\frac{m^2}{\eta^2}\right)F(\mathbf{p}|\eta,\eta) \right].
\end{equation}
Now, considering the terms, which arise from (\ref{eq:Bubble_contribution}) when time derivatives act on the theta-functions of the propagators (\ref{eq:dS_propagators}), we obtain:
\beq \label{eq:locals_from_bubble}
\begin{aligned}
    \Delta \Pi^{\text{loc}}_{00|00}(\eta) &= \frac{1}{2}\frac{1}{H^2\eta^2}\int_{p}\partial_{\eta}\partial_{\eta'}F\bigg|_{\eta'=\eta}, \quad
    \Delta \Pi^{\text{loc}}_{00|0k}(\eta) = \Delta \Pi^{\text{loc}}_{0i|kl}(\eta) = 0, 
    \\
    \Delta \Pi^{\text{loc}}_{0i|0k}(\eta) &= \frac{1}{H^2\eta^2}\frac{1}{2(D-1)}\delta_{ik}\int_{p}p^2 F(\mathbf{p}|\eta,\eta),\quad
    \Delta \Pi^{\text{loc}}_{00|kl}(\eta) = \frac{1}{2}\frac{1}{H^2\eta^2}\delta_{kl}\int_{p}\partial_{\eta}\partial_{\eta'}F\bigg|_{\eta'=\eta},
    \\
    \Delta \Pi^{\text{loc}}_{ij|kl}(\eta) &= \frac{1}{2}\frac{1}{H^2\eta^2}\delta_{ij}\delta_{kl}\int_{p}\partial_{\eta}\partial_{\eta'}F\bigg|_{\eta'=\eta}.
\end{aligned}
\eeq
Finally, summing up all the local contributions (\ref{eq:tadpole}) and (\ref{eq:locals_from_bubble}) we find from (\ref{eq:Local_contribution}):
\beq \label{eq:local_polarization_operator}
\begin{aligned}
    \Pi^{\text{loc}}_{00|00}(\eta) &= \frac{1}{H^2\eta^2}\frac{1}{4}\left\langle T_{00}^{cl-cl} \right\rangle, \quad
    \Pi^{\text{loc}}_{ij|00}(\eta) = -\frac{1}{H^2\eta^2}\frac{1}{4}\left\langle T_{ij}^{cl-cl} \right\rangle,\quad
    \Pi^{\text{loc}}_{0i|0l}(\eta) = -\frac{1}{H^2\eta^2}\frac{1}{4}\left\langle T_{00}^{cl-cl} \right\rangle\delta_{il}, 
    \\
    \Pi^{\text{loc}}_{ij|kl}(\eta) &= \frac{1}{H^2\eta^2}\left\{\pi_{3}(\eta)\left( \delta_{ij}\delta_{kl} - \delta_{ik}\delta_{jl} - \delta_{il}\delta_{jk} \right) - \frac{1}{4}\int_{p}\partial_{\eta}\partial_{\eta'}F\bigg|_{\eta'=\eta}\delta_{ij}\delta_{kl}   \right\}.
\end{aligned}
\eeq
Having the explicit expressions for all important parts of the effective action, we can write it down as follows (we omit the Einstein-Hilbert part):
\begin{gather}
    \Gamma_{\text{eff}} = -\frac{1}{2}\int \frac{d^Dx}{H^D\eta^D}h_{q}^{\alpha\beta}\left\langle T_{\alpha\beta}^{cl-cl} \right\rangle - \frac{1}{2}\int_0^{\infty}\frac{d\eta}{H^D\eta^D}\int_{\eta}^{\infty}\frac{d\eta'}{H^D\eta'^{D}}\int_{k}h_{cl}^{\mu\nu}(\eta',\mathbf{k})\Pi^{\text{bub}}_{\mu\nu|\alpha\beta}(-\mathbf{k}|\eta',\eta)h_{q}^{\alpha\beta}(\eta,-\mathbf{k}) + \nonumber
    \\
    + \int_{0}^{\infty}\frac{d\eta}{H^D\eta^D}\int_{k} h_{cl}^{\mu\nu}(\eta,\mathbf{k})\Pi^{\text{loc}}_{\mu\nu|\alpha\beta}(\eta)h_{q}^{\alpha\beta}(\eta,-\mathbf{k}).
\label{eq:effective_action}
\end{gather}
Let us emphasize at this point that all the local contributions (\ref{eq:Local_contribution}) and averaged lagrangian (\ref{eq:lagrangian_averaged}) contain the Keldysh function at coincident points, which is the UV-divergent quantity and requires an accurate regularization procedure, which must preserve the symmetries of the theory. Nevertheless, these terms are indispensable for gauge invariance. Indeed, one can check the gauge symmetry of this action, using the transformation in the zeroth and first orders in perturbation:
\begin{equation}\label{eq:gauge_transformation}
    \delta_{\xi}h^{\mu\nu} = -\hat{g}^{\mu\lambda}\hat{\nabla}_{\lambda}\xi^{\nu} - h^{\nu\beta}\hat{g}_{\alpha\beta}\hat{g}^{\mu\lambda}\hat{\nabla}_{\lambda}\xi^{\alpha} - \hat{g}^{\mu\lambda}\Gamma^{(1)\nu}_{\lambda\omega}\xi^{\omega} + \big\{\mu\leftrightarrow \nu\big\},
\end{equation}
where $\Gamma^{(1)\nu}_{\lambda\omega}$ are the first order corrections to the exact Christoffel symbols in the metric (\ref{eq:perturbed_dS}). The invariance in the order $\mathcal{O}(\xi)$ is guaranteed by the covariant conservation of the stress-energy tensor. In Appendix \ref{App:Gauge_inv} we show how to make sure of gauge invariance in the order $\mathcal{O}\left(||h_q\cdot\xi||\right)$ with the expressions for the polarization operators given in this section.

In the case of BD-state we must have $\left\langle T_{\mu\nu}^{cl-cl} \right\rangle = \delta\lambda \hat{g}_{\mu\nu}$\footnote{This statement is not trivial and can be seen explicitly only in the regularization schemes which preserves dS isometries, such as dimensional regularization or point-splitting method \cite{Prokopec:2002uw, 1977RSPSA.354...59D, PhysRevD.14.2490, Bunch:1978yq}. It is a separate interesting topic, that even in the thermal state the situation is much more subtle for the space-times with horizons \cite{Bazarov:2021rrb}.}, so that the first ``source''--term in (\ref{eq:effective_action}) is attributed to the renormalization of the cosmological constant: $\Lambda_{\text{ren}} = \Lambda + 8\pi G\delta\lambda$. More accurately, let us subtract the following $\Lambda$-renormalization counterterm from the one-loop answer (\ref{eq:effective_action}), which we also write in terms of the fields (\ref{eq:Keldysh_rotation}) after Keldysh rotation:
\begin{multline}\label{eq:Lambda_counterterm}
     \delta_{\Lambda}\Gamma_{\text{eff}} = -\int d^Dx\sqrt{g(x)}\delta\lambda = -\int d^Dx\sqrt{\hat{g}}\delta\lambda - \frac{1}{2} \int d^Dx\sqrt{\hat{g}(x)}\hat{g}_{\mu\nu}h^{\mu\nu}_{q}\times \delta\lambda- \\
     - \frac{1}{4}\int d^Dx\sqrt{\hat{g}}\left(\hat{g}_{\mu\nu} \hat{g}_{\alpha\beta} - \hat{g}_{\mu\alpha} \hat{g}_{\nu\beta} - \hat{g}_{\mu\beta} \hat{g}_{\nu\alpha}\right)h^{\mu\nu}_{cl}h^{q}_{\alpha\beta}\times\delta\lambda.
\end{multline}
As we see, this renormalization affects only the local contributions from the loops and, if we set
\begin{equation}
    \delta\lambda = -\frac{1}{2}\left\langle \mathscr{L}\right\rangle - H^2\eta^2\frac{1}{D-1}\int_{p}p^2F(\mathbf{p}|\eta,\eta),
\end{equation}
it eliminates the ``source''-term and the most of the local parts (\ref{eq:local_polarization_operator}):
\begin{equation}
    \begin{aligned}
        \widetilde{\Pi}^{\text{loc}}_{00|00}(\eta) &= 0, \quad\widetilde{\Pi}^{\text{loc}}_{ij|00}(\eta) = 0, \quad\widetilde{\Pi}^{\text{loc}}_{0i|0l}(\eta) = 0,\quad\quad\quad\quad\quad 
        \\
        \widetilde{\Pi}^{\text{loc}}_{ij|kl}(\eta) &= \frac{1}{H^2\eta^2}\left\{\widetilde{\pi}_{3}(\eta)\left( \delta_{ij}\delta_{kl} - \delta_{ik}\delta_{jl} - \delta_{il}\delta_{jk} \right) - \frac{1}{4}\int_{p}\partial_{\eta}\partial_{\eta'}F\bigg|_{\eta'=\eta}\delta_{ij}\delta_{kl}   \right\}, 
    \end{aligned}
\end{equation}
where
\begin{equation}\label{eq:renormalized_pi3}
    \widetilde{\pi}_{3} = \frac{1}{4}\frac{1}{D-1}\int_{p}p^2 F(\mathbf{p}|\eta,\eta).
\end{equation}
\section{Effective mass of the tensor mode}\label{sec:Tensor_mode}
Having the expressions for the quantum corrections to the induced gravity action in terms of specific integrals, we can investigate the effective equation of motion in detail. In the case of the tensor sector $\mathscr{h}_{00} = \mathscr{h}_{0i} = 0, \; \mathscr{h}_{ij} = \mathscr{h}_{ij}^{TT}$, the only non-vanishing component of the eq. (\ref{eq:Eff_EOM}) reads (see Appendix \ref{subSec:Bubble} for the notations in the non-local part):
\begin{equation}\label{eq:almost_equation_for_htt}
    \nabla_{\eta}\partial_{\eta}\mathscr{h}_{ij}^{TT} + k^2 \mathscr{h}_{ij}^{TT} - 32\pi G\int_{\eta}^{\infty}\frac{d\eta'}{H^{D-2}\eta'^{D-2}}e_5(\mathbf{k}|\eta,\eta')\mathscr{h}_{ij}^{TT}(\mathbf{k},\eta') + 64\pi G\widetilde{\pi}_{3}(\eta)\mathscr{h}_{ij}^{TT}(\mathbf{k},\eta) = 0.
\end{equation}
We see that (\ref{eq:almost_equation_for_htt}) is an integro-differential equation, so the notion of mass requires accuracy. Following the approach of \cite{Boyanovsky:1999jh}, where the effective mass of photon in the systems out of the thermal equilibrium was introduced, we expand the integral-part of the eq. (\ref{eq:almost_equation_for_htt}) in derivatives of $\mathscr{h}_{ij}^{TT}$ in time. Namely, if we denote 
\beq\label{eq:expanding_bubble}
\begin{aligned}
\Gamma^{\text{bub}}\left(\eta,\eta'\right) &= -32\pi G \int_{\eta'}^{\infty}\frac{d\eta''}{H^{D-2}\eta''^{D-2}}e_5(\mathbf{k}|\eta,\eta''), 
\\
\partial_{\eta'}\Gamma^{\text{bub}}\left(\eta,\eta'\right) &= 32\pi G \frac{1}{H^{D-2}\eta'^{D-2}}e_5(\mathbf{k}|\eta,\eta')
\end{aligned}
\eeq
and then integrate (\ref{eq:almost_equation_for_htt}) by parts, we arrive at
\begin{equation}\label{eq:equation_for_htt}
\begin{aligned}
    \bigg\{\nabla_{\eta}\partial_{\eta} + k^2 - 32\pi G\int_{\eta}^{\infty}\frac{d\eta'}{H^{D-2}\eta'^{D-2}}e_5(\mathbf{k}|\eta,\eta') + 64\pi G\widetilde{\pi}_{3}(\eta)\bigg\}\mathscr{h}_{ij}^{TT}(\mathbf{k},\eta) - \\
    + \int_{\eta}^{\infty} d\eta'\Gamma^{\text{bub}}\left(\eta,\eta'\right)\partial_{\eta'}\mathscr{h}_{ij}^{TT}(\mathbf{k},\eta') = 0.
\end{aligned}
\end{equation}
One can continue this procedure and expand the non-local part of the effective action through multiple time derivatives of $\mathscr{h}_{ij}^{TT}(\mathbf{k},\eta)$. Then, for slowly varying field $\mathscr{h}_{ij}^{TT}(\mathbf{k},\eta)$ one has the Klein-Gordon equation of type (\ref{eq:EOM_for_harmonics}):
\begin{equation}\label{eq:KG_equation_for_htt}
    \nabla_{\eta}\partial_{\eta}\mathscr{h}_{ij}^{TT} + \left[k^2 + \frac{m_{TT}^{2}\left(k,\eta\right)}{\eta^2}\right]\mathscr{h}_{ij}^{TT} \simeq 0.
\end{equation}
The last equation allows us to define an effective mass for graviton as one does for non-equilibrium systems \cite{Boyanovsky:1999jh}:
\begin{equation}\label{eq:mass_of_htt_def}
    m_{TT}^{2} = -32\pi G\lim_{k\rightarrow 0}\lim_{\eta \rightarrow 0}\eta^2\times\left[ \int_{\eta}^{\infty} \frac{d\eta'}{H^{D-2}\eta'^{D-2}}e_5\left(\mathbf{k}|\eta,\eta'\right)  - 2\widetilde{\pi}_{3}\left(\eta\right)\right].
\end{equation}
In general situation, the order of the limits in (\ref{eq:mass_of_htt_def}) is very important. In particular, in flat space another order leads to the immediate zero value for the Debye mass \cite{Popov:2017xut, Polyakov:2022dpa}. We define the limits in the way they are commonly taken in condensed matter physics \cite{Boyanovsky:1999jh, tsvelik_2003}, where this order is also physically approved. However, it can be easily seen that in our case the quantity $m^2_{TT}\left(k,\eta\right)$ actually depends on the dS invariant variable $k\eta$, so that the only limit we need to take is the zero limit for physical momentum $k\eta \rightarrow 0$. 

At first glance it may seem that the integration over time region from $\eta$ to $\infty$ may bring some infra--red effects to the mass $m^2_{TT}$ and the local correction $\widetilde{\pi}_3$ just removes some ultraviolet singularities. However, the quantum mechanical perturbation theory (see \cite{Popov:2017xut} and Appendix \ref{App:Keldysh_function}) provides us with the formula
\begin{equation}\label{eq:relation_for_Keldysh}
    \partial_{p^2}F(\mathbf{p}|\eta,\eta) = -2\int_{\eta}^{\infty}\frac{d\eta'}{H^{D-2}\eta'^{D-2}}F(\mathbf{p}|\eta,\eta')\rho(\mathbf{p}|\eta,\eta'),
\end{equation}
which reduces the first term on the RHS of (\ref{eq:mass_of_htt_def}) to the similar local contribution as the second one. Now we use that $\partial_{p^2}F(\mathbf{p}|\eta,\eta) = \frac{1}{2p}\partial_{p}F$ and directly find in the limit $k\rightarrow 0$:
\beq \label{eq:calculation_of_htt_mass}  
\begin{aligned}
    \int_{\eta}^{\infty} \frac{d\eta'}{H^{D-2}\eta'^{D-2}}e_5\left(\mathbf{k}|\eta,\eta'\right) =
    \\
    =\frac{2}{D(D-2)}\int_{\eta}^{\infty} \frac{d\eta'}{H^{D-2}\eta'^{D-2}}\int_{p} \left(p^2 -\frac{(\mathbf{kp})^2}{k^2}\right)^2 F(\mathbf{p}|\eta,\eta')\rho(\mathbf{k}-\mathbf{p}|\eta,\eta')= 
    \\
    \overset{k\rightarrow 0}{=} -\frac{1}{D(D-2)}\int_{p} \left(p^2 -\frac{(\mathbf{kp})^2}{k^2}\right)^2 \partial_{p^2}F(\mathbf{p}|\eta,\eta) = 
    \\
    = -\frac{1}{D(D-2)}\frac{\Omega_{d}}{(2\pi)^d}\int_{0}^{\infty}dp p^{D+2}\left[1 - \frac{2}{D-1} + \frac{3}{D^2-1}\right] \frac{1}{2p}\partial_{p}F = 
    \\
    = \frac{1}{2}\frac{\Omega_{d}}{(2\pi)^d(D-1)}\int_{0}^{\infty}dp p^{D}F(\mathbf{p}|\eta,\eta) = \frac{1}{2}\frac{1}{D-1}\int_{p}p^2F(\mathbf{p}|\eta,\eta),  
\end{aligned} 
\eeq
where in the third line we have integrated over the angles and then by parts over the absolute value of the momentum. Eventually, we take the renormalized value $\widetilde{\pi}_3$ (\ref{eq:renormalized_pi3}) and find from the definition (\ref{eq:mass_of_htt_def}) that the mass of the spin-2 metric perturbation vanishes:
\begin{equation}\label{eq:mass_of_htt}
    m_{TT}^{2} = 0.
\end{equation}

As it was noted in the Introduction, we believe that there is a special reason why we have no mass generation for photon \cite{Popov:2017xut} and graviton in dS, while it was proved that there can be mass of the spin-$2$ graviton \cite{Porrati:2001db, Porrati:2003sa} in $\text{AdS}_4$\footnote{Strictly speaking, quantum field theory in global AdS is ill defined and suffers from unusual ultraviolet phenomena \cite{Akhmedov:2018lkp, Akhmedov:2020jsi}. However, it serves us with a useful playing background to investigate properties of QFT in different space-times with high number of symmetries.}. Namely, the mass of the gauge fields generates if there is a pole in the non-local part of the self-energy appears, which corresponds to the Goldstone boson as on the fig.\ref{fig:Higgs}. The necessary condition for this is the presence of this Goldstone boson in the tensor product of the from $D(E_1,0)\otimes D(E_2,0)$, where $D(E,0)$ is an infinite-dimensional, irreducible, positive-weight representation of the isometry group of the embedding space (UIR), which corresponds to the physical states of the scalar field theory. Here $E$ and $s$ correspond to the minimal energy and angular momentum (spin) of the given representation, such that other states in it are obtained by the action of the appropriate creation operators \cite{Nicolai:1984hb}. In the case of $\text{AdS}_4$ the isometry group is $SO(2,3)$ and we have the relations \cite{Porrati:2001db, Fronsdal:1978vb}:
\begin{equation}\label{eq:nearly_massless_decomposition_of_UIR_AdS}
    D(E,s) \rightarrow D(s+1,s) \oplus D(s+2, s-1) \;\text{as }E\rightarrow s+1,
\end{equation}
\begin{equation}\label{eq:tensor_product_of_UIRs_in_AdS}
    D(E_1,0)\otimes D(E_2,0) = \sum\limits_{l=0}^{\infty}\sum\limits_{n=0}^{\infty}D(E_1+E_2+l+2n, l).
\end{equation}
The first line (\ref{eq:nearly_massless_decomposition_of_UIR_AdS}) shows that the field of spin-$s$ in the massless limit $E\rightarrow s+1$ decomposes onto the massless field from $D(s+1,s)$ of the same spin and the field of spin-$(s-1)$. This decomposition tells us that the field becomes massive after swallowing 
 the boson from the representation $D(s+2,s-1)$ in $\text{AdS}_4$. In the cases of photon ($s=1$) and graviton ($s=2$) the corresponding Goldstone bosons are from $D(3,0)$ and $D(4,1)$. The second relation (\ref{eq:tensor_product_of_UIRs_in_AdS}) shows that the states of these gauge bosons may appear in the non-local contribution to the self-energy of either photon or graviton in the case of conformally coupled scalar, which corresponds to the choice $E=1$ or $E=2$. In the works \cite{Porrati:2001db, Porrati:2003sa} it was shown that the Goldstone vector is indeed present in the graviton's self energy for certain boundary conditions, hence the mass of the graviton generates in $\text{AdS}_4$. In order to complete the considerations of the mass of the photon in \cite{Popov:2017xut} from this point of view, we investigate the scalar QED in $\text{AdS}_4$ with curvature radius $L$ in Appendix \ref{App:Photon_AdS}. 
 
 In contrast to AdS, there are no such Goldstone bosons in the tensor product of UIRs \cite{Penedones:2023uqc} for dS isometry group, hence the absence of the photon's and graviton's mass is expected. However, dS is neither stationary nor stable background, so it would be rather naive to proceed this way and one'd better adopt the non-equilibrium approach, that we use in this paper. Moreover, it is argued in many works \cite{Akhmedov:2011pj,Anderson:2013zia, Akhmedov:2013vka, Prokopec:2011ms, Serreau:2013psa} that there are a lot of IR-peculiarities in loop corrections in dS, which may affect the result significantly. 
\section{Discussion on the scalar sector of gravity}\label{sec:Scalar_sector}
\subsection{General remarks}\label{subSec:General_remarks_on_scalar_sector}
Treating the problem in transverse-traceless gauge, we apparently can get some information only about the coefficient $A\left(\Delta_L\right)$ in (\ref{eq:Eff_EOM_dS_inv}). In order to say something about $E\left(\Delta_L\right)$ one has to include into consideration different modes of metric's perturbation (\ref{eq:spiral_decomposition}). For instance, if we naively set $g_{\mu\nu} = e^{2\sigma}\hat{g}_{\mu\nu}$ we obtain from (\ref{eq:Eff_EOM_dS_inv}):
\beq\label{eq:eff_action_for_conf_factor}
   \Gamma_{\text{eff}} \propto \int\left[\sigma\left(\Delta_L + DH^2\right)\sigma + 32\pi G\frac{D-1}{D-2}\times\sigma\cdot\left(\frac{\Delta_L + DH^2}{\Delta_L + \left(D-1\right)H^2}\right)^2 E\left(\Delta_L\right)\cdot\sigma\right].
\eeq
Let us make a few observations. First, we see that $E\left(\Delta_L\right)$ determines a shift to ``mass'' $DH^2$ which conformal parameter $\sigma$ already has. Second, it is crucial that both the kinetic and mass terms enter the effective action with the ghost-like sign. On the classical level in the presence of classical matter this leads to Jeans' instability \cite{Gorbunov:1354521, gorbunov2011introduction} -- it is not surprising, however, that the classical equations of motion don't have non-trivial solutions without matter (the other constraints of (\ref{eq:Linearized_Einstein_EOM}) are not satisfied): in this case the field $\sigma$ is non-propagating. On the other hand, nobody exactly knows what happens in loop-modified gravity, because, e.g. in the naive massive gravity \cite{Jaccard:2012ut}, the scalar ghost-like degrees of freedom become dynamical. To answer these questions, more thorough investigation of $\Gamma_{\text{eff}}$ is necessary. 

To find the terms which contribute to $E\left(\Delta_L\right)$, it is convenient to express the bubble diagram in terms of commutator of stress-energy tensors in the operator formalism:
\beq\label{eq:bubble_as_commutator}
    \Pi_{\mu\nu|\alpha\beta}^{\text{bub}}\left(\mathbf{k}|\eta,\eta'\right) = -\frac{1}{8}\bigg\langle 
 \text{BD}\bigg|\big[T_{\mu\nu}\left(\mathbf{k},\eta\right), T_{\alpha\beta}\left(-\mathbf{k},\eta'\right)\big]\bigg|\text{BD}\bigg\rangle.
\eeq
Then, using the relations derived in Appendix \ref{App:Relations_for_STE}, we can write for the correction to $\sigma$'s mass-like term, which is given by the trace of self-energy over $\mu=\nu$ and $\alpha=\beta$ indices (before taking the limit $k\eta\rightarrow 0$):
\beq\label{eq:sigma_mass_correction}
    \delta m_{\sigma}^2 \propto \int_{\eta}^{\infty} \frac{d\eta'}{H^{D-2}\eta'^{D-2}}\left[D\left(D-2\right)\Pi_{00|00}\left(\mathbf{k}|\eta,\eta'\right) + \Pi_{\Delta T|\Delta T}\left(\mathbf{k}|\eta,\eta'\right)\right] - 2\Pi_{ii|kk}^{\text{loc}},
\eeq
where $\Pi_{\Delta T|\Delta T}$ denotes the commutator of the form (\ref{eq:bubble_as_commutator}) for $\Delta T$ defined in Appendix \ref{eq:Trace_stress_tensor}. The quantity (\ref{eq:sigma_mass_correction}) is UV-divergent even in flat space, not mentioning the problems with proper renormalization in dS\footnote{Actually, being a short-distance local phenomenon, UV renormalization must be the same for any gravitational background at least at the leading order. Nevertheless, in order to obtain correct values for IR quantities, one should preserve symmetries of the theory at each step of the calculation, which, for example, forbids the naive UV cut-off regularization scheme in de Sitter.}\cite{tHooft:1974toh, Frob:2013sxa, Park:2011ww, Ford:2004wc}. This is not the end of the story. After the accurate subtractions of required counterterms, we still may have some ``spurious'' divergences left in the quantity $\delta m_{\sigma}^2$ as a consequence of definition of the induced mass as a coefficient in the expansion of the effective action in time derivatives of $\mathscr{h}_{\mu\nu}$. This is similar to the Taylor expansion of the function $e^{-x^2} = 1-x^2+\ldots$, where the whole function is convergent in the limit $x\rightarrow \infty$, while each term in the expansion is divergent. The appearance of such peculiarities can be seen if one considers closely the first term in (\ref{eq:sigma_mass_correction}) in the limit $k\rightarrow 0$ in $\mathbf{x}$-space:
\beq\label{eq:Pi_OO_as_commutator}
\begin{aligned}
    D\left(D-2\right)\int_{\eta}^{\infty} \frac{d\eta'}{H^{D-2}\eta'^{D-2}}\Pi_{00|00}\left(\mathbf{0}|\eta,\eta'\right) \propto
    \\
    \propto \int_{\eta}^{\infty}\frac{d\eta'}{\eta'^{D-2}}\left\langle BD\bigg|\left[T_{00}\left(\mathbf{0}, \eta\right), \int_{\mathbf{x}} T_{00}\left(\mathbf{x}, \eta'\right)\right] \bigg|BD\right\rangle.
\end{aligned}
\eeq
Normally, the integral of $T_{00}$ is the conserved charge (total energy), which commutes with any operator. For example, the covariant conservation of the electrical current $\hat{\nabla}_{\mu}J^{\mu} = \partial_{\eta}J_{0} + \frac{D}{\eta}J^{0} - \partial_{i}J_{i} = 0$ implies that the integral of $J_{0}$ over the position space is $J_{0} = \frac{1}{\eta^D}\times\text{const}$. Hence, in any commutator of the form (\ref{eq:Pi_OO_as_commutator}) with electrical charge the dependence of the charge on time factors out and the result is exactly  zero, which leads to the vanishing Debye mass in dS \cite{Popov:2017xut}. In contrast, the covariant conservation condition for the stress-energy tensor includes additional term $\Delta T$ (\ref{eq:conserv_stress_tensor}), which makes the analogy with the electric charge inapplicable. This conclusion establishes the fact that the energy (at least defined as the integral of $T_{00}$) isn't conserved in non-stationary background and non-trivially commutes with other operators. Therefore, although in flat space one has identically zero contribution from $\Pi_{00|00}$ to the $\delta m_{\sigma}^2$, in dS we obtain:
\beq
\begin{aligned}
    \delta m_{\sigma}^2 \sim \int_{1}^{\infty} \frac{d\tau}{\tau}\int \frac{d^{D-1}\boldsymbol{\xi}}{\left(2\pi\right)^{D-1}}\text{Im}\left\{ t_{00}\left(\xi\right)t_{00}\left(\xi\tau\right) \right\} + \ldots,
    \\
    t_{00}\left(\xi\right) \equiv \left(\frac{D-1}{2}h_{\nu}(\xi) + \xi h_{\nu}'(\xi)\right)^2 + \left(\xi^2 + m^2\right)h_{\nu}^2(\xi),
    \\
    \tau = \frac{\eta'}{\eta}, \;\boldsymbol{\xi} = \eta\mathbf{p}.
\end{aligned}
\eeq
Taking the mode functions (\ref{eq:harmonics}), one can verify that in $D>2$ the are divergences in the UV region. Therefore, in view of the fact that these divergences are not universal, we believe that they contribute to some well-defined parts of $\Gamma_{\text{eff}}$ after the resummation, but in this case we should adopt more accurate approaches, such as K\"all\'en-Lehmann decomposition for the correlators in dS \cite{Loparco:2023rug}. We leave the treatment of the issues discussed in this subsection for future work, and below we investigate the simplest case of two-dimensional space-time.
\subsection{Two-dimensional space-time}\label{subSec:2D}
\begin{figure}[hbt!]
    \centering
\begingroup%
  \makeatletter%
  \providecommand\color[2][]{%
    \errmessage{(Inkscape) Color is used for the text in Inkscape, but the package 'color.sty' is not loaded}%
    \renewcommand\color[2][]{}%
  }%
  \providecommand\transparent[1]{%
    \errmessage{(Inkscape) Transparency is used (non-zero) for the text in Inkscape, but the package 'transparent.sty' is not loaded}%
    \renewcommand\transparent[1]{}%
  }%
  \providecommand\rotatebox[2]{#2}%
  \newcommand*\fsize{\dimexpr\f@size pt\relax}%
  \newcommand*\lineheight[1]{\fontsize{\fsize}{#1\fsize}\selectfont}%
  \ifx\svgwidth\undefined%
    \setlength{\unitlength}{475.27578212bp}%
    \ifx\svgscale\undefined%
      \relax%
    \else%
      \setlength{\unitlength}{\unitlength * \real{\svgscale}}%
    \fi%
  \else%
    \setlength{\unitlength}{\svgwidth}%
  \fi%
  \global\let\svgwidth\undefined%
  \global\let\svgscale\undefined%
  \makeatother%
  \begin{picture}(1,0.43520099)%
    \lineheight{1}%
    \setlength\tabcolsep{0pt}%
    \put(0,0){\includegraphics[width=\unitlength,page=1]{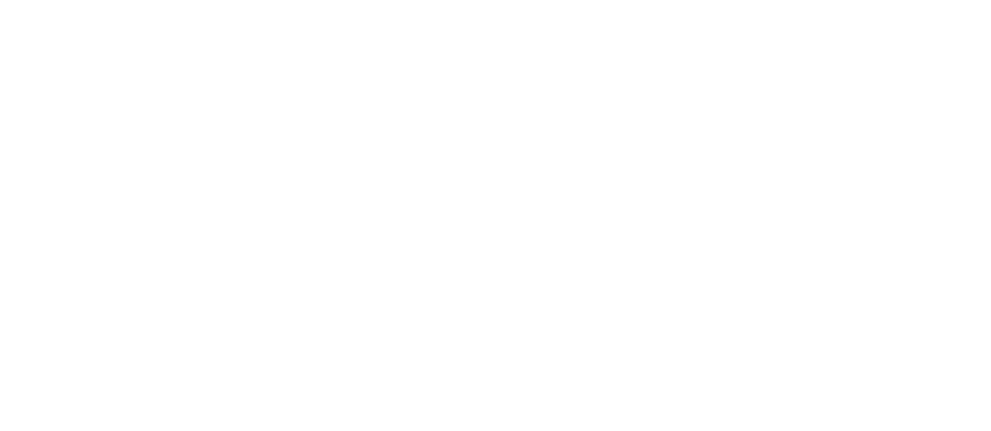}}%
    \put(0.82112613,0.02191076){\color[rgb]{0.0745098,0.03529412,0.03529412}\makebox(0,0)[lt]{\lineheight{1.25}\smash{\begin{tabular}[t]{l}$m^2$\end{tabular}}}}%
    \put(0,0){\includegraphics[width=\unitlength,page=2]{2dPlot_DS+MINK.pdf}}%
    \put(0.1815591,0.40447509){\color[rgb]{0.0745098,0.03529412,0.03529412}\makebox(0,0)[lt]{\lineheight{1.25}\smash{\begin{tabular}[t]{l}$m^2_{\sigma}$\end{tabular}}}}%
  \end{picture}%
\endgroup%

    \caption{The dependence of $m^2_{\sigma}$ on the squared mass of the scalar field. The orange line depicts the value in flat space-time. All masses are measured in the units of Hubble parameter.}
    \label{fig:2d_Plot}
\end{figure}

It is well-known that in $2D$ the Einstein-Hilbert action is topological and the only independent component of metric's perturbation is the Weyl parameter $\sigma$. Moreover, the kinetic term comes form the loops and also has ghost like sign in the effective action, if we take into account only the matter field with positive central charge $c$ \cite{Polyakov:1981rd, Erbin2015NotesO2} -- below we insert the mass term to the action with ghost-like sign as well, such as it appears in the KG equation of motion for $\sigma$ as a standard mass, assuming $c>0$. In addition, it can be seen that the tadpole diagram in $2D$ contributes only to the cosmological constant's renormalization, and the loop diagram is given by the commutator of the form (\ref{eq:bubble_as_commutator}) of two stress-energy tensor's traces. Hence, because the only covariant quantities which constitute to the effective action is the covariant laplacian $\Box$ and Ricci scalar $R$ in two dimensions, when we immerse the massive scalar field in the curved background, we expect $\Gamma_{\text{eff}}$ to have the following form (in the second order in $\sigma$):
\beq\label{eq:effAction_2D}
    \Gamma_{\text{eff}} = \int\left[\delta m_{\sigma}^2 R\frac{1}{\Box^2}R + \frac{c}{96\pi}R\frac{1}{\Box}R + \ldots\right],
\eeq
where the ellipsis stand, first, for further expansion of the effective action in the powers of laplacian and, second, for less trivial terms such as Mabuchi action \cite{Ferrari:2011rk, delacroixdelavalette:tel-01706737, Bilal:2017zhb, Bilal:2021fqq}, which arises as a modification of the Liouville action for non-conformal matter interacting with the metric on a Riemannian manifold\footnote{Although Mabuchi action is well-defined on Riemann surfaces of fixed area with boundary, appearance of its parts in the induced gravity seems to be universal.}. This action satisfies cocycle condition, is bounded from below, and affects the calculation of correlators in modified two-dimensional quantum gravity \cite{Bilal:2013ska}. Furthermore, it has a natural generalization to higher dimensions, making the study of this contribution a separate interesting task. We write $\delta m_{\sigma}^2$ in (\ref{eq:effAction_2D}) instead of $m_{\sigma}^2$, because such terms as non-perturbative Mabuchi action certainly lead to contributions to the quadratic part of $\Gamma_{\text{eff}}$ being formally expanded in $\sigma$ \cite{delacroixdelavalette:tel-01706737, Bilal:2017zhb}:
\beq\label{eq:Mabuchi_action}
\begin{aligned}
    S_{\text{Mabuchi}}\left[g_{\mu\nu}, \hat{g}_{\mu\nu}\right] &= \int\sqrt{|\hat{g}|}\left( 
\frac{1}{\pi A}\sigma e^{2\sigma} +\ldots\right),
\\
S_{\text{grav}} \left[g_{\mu\nu}, \hat{g}_{\mu\nu}\right] &\equiv \frac{1}{2}\log\frac{\text{det}\big(-\Box + M^2\big)}{\text{det}\big(-\hat{\Box}+M^2\big)} = \\
 &=\frac{c}{96\pi}\int R\frac{1}{\Box}R + \frac{M^2A}{4}S_{\text{Mabuchi}}\left[g_{\mu\nu}, \hat{g}_{\mu\nu}\right] + \mathcal{O}\left(M^4\right),
\end{aligned}
\eeq
where $A$ is the area of the Riemann surface (here we assume the euclidean signature and zero genus), on which this action is defined. Indeed, for the definition of $m_{\sigma}^2$ as in the previous section, we find in flat space for the plane-wave harmonics $f_{\mathbf{p}}(t) = \frac{1}{\sqrt{2}\sqrt[4]{p^2+M^2}}e^{-i\sqrt{p^2+M^2}t}$ (we measure mass in the units of $H$ here):
\beq\label{eq:sigma_2d_flat}
    m_{\sigma}^2 = \frac{1}{H^2}\frac{4M^4}{\pi}\int_{-\infty}^{t}dt'\int_{0}^{\infty}dp\;\text{Im}\left\{f_{\mathbf{p}}(t')f_{\mathbf{p}}(t')f_{\mathbf{p}}^*(t)f_{\mathbf{p}}^*(t)  \right\} = \frac{m^2}{2\pi}.
\eeq
For the dS background we have (here $h(\xi)$ denotes the harmonic function (\ref{eq:harmonics}) eather for complementary or for principal series):
\beq\label{eq:sigma_dS_2D}
    m_{\sigma}^2 = -\frac{4 m^4}{\pi}\int_{1}^{\infty}\frac{d\tau}{\tau}\int_{0}^{\infty}d\xi\;\text{Im}\left\{ h^{2}\left(\xi\right)h^{*2}\left(\xi\tau\right) \right\}.
\eeq
This integral is convergent and can be evaluated numerically (see fig.\ref{fig:2d_Plot}). It is interesting, that the result in dS considerably deviates for small masses of scalar field, while the answer for the static Riemann manifold (\ref{eq:Mabuchi_action}) is supposed to be the leading contribution in this region. This makes us believe that the IR behaviour of light scalar field in dS leads to amplification of $\delta m_{\sigma}^2$ in the long-wave expansion of the effective action (\ref{eq:effAction_2D}). Meanwhile, we see on the fig.\ref{fig:2d_Plot} that for large mass of the scalar field the value of $m^2_{\sigma}$ approaches the flat-space value, which is not surprising: very heavy fields decouple and don't feel the effects of the background. Let us confirm this analytically for fermionic matter.
\subsubsection*{Fermionic fields in 2D}
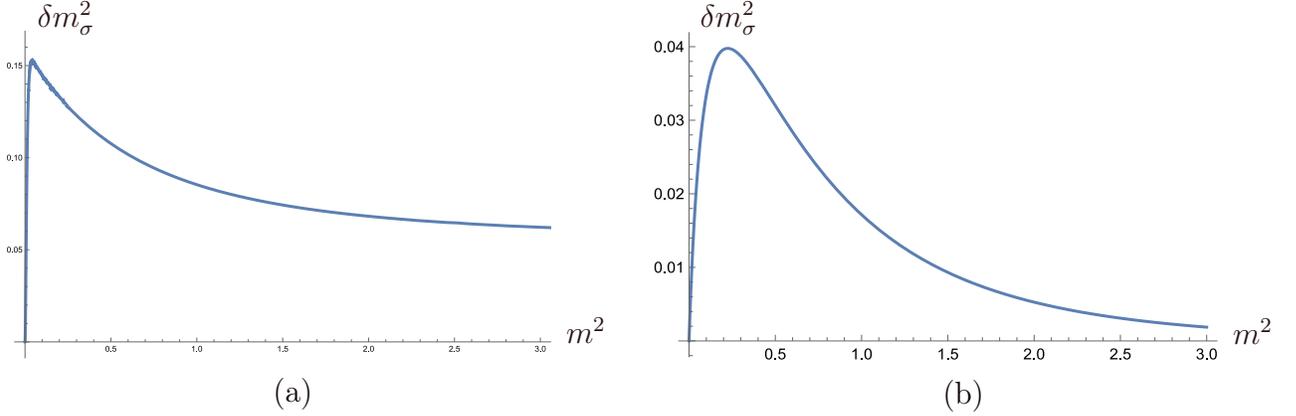
\begin{figure}[hbt!]
    \centering
\begingroup%
  \makeatletter%
  \providecommand\color[2][]{%
    \errmessage{(Inkscape) Color is used for the text in Inkscape, but the package 'color.sty' is not loaded}%
    \renewcommand\color[2][]{}%
  }%
  \providecommand\transparent[1]{%
    \errmessage{(Inkscape) Transparency is used (non-zero) for the text in Inkscape, but the package 'transparent.sty' is not loaded}%
    \renewcommand\transparent[1]{}%
  }%
  \providecommand\rotatebox[2]{#2}%
  \newcommand*\fsize{\dimexpr\f@size pt\relax}%
  \newcommand*\lineheight[1]{\fontsize{\fsize}{#1\fsize}\selectfont}%
  \ifx\svgwidth\undefined%
    \setlength{\unitlength}{516.40643065bp}%
    \ifx\svgscale\undefined%
      \relax%
    \else%
      \setlength{\unitlength}{\unitlength * \real{\svgscale}}%
    \fi%
  \else%
    \setlength{\unitlength}{\svgwidth}%
  \fi%
  \global\let\svgwidth\undefined%
  \global\let\svgscale\undefined%
  \makeatother%
  \begin{picture}(1,0.34513277)%
    \lineheight{1}%
    \setlength\tabcolsep{0pt}%
    \put(0,0){\includegraphics[width=\unitlength,page=1]{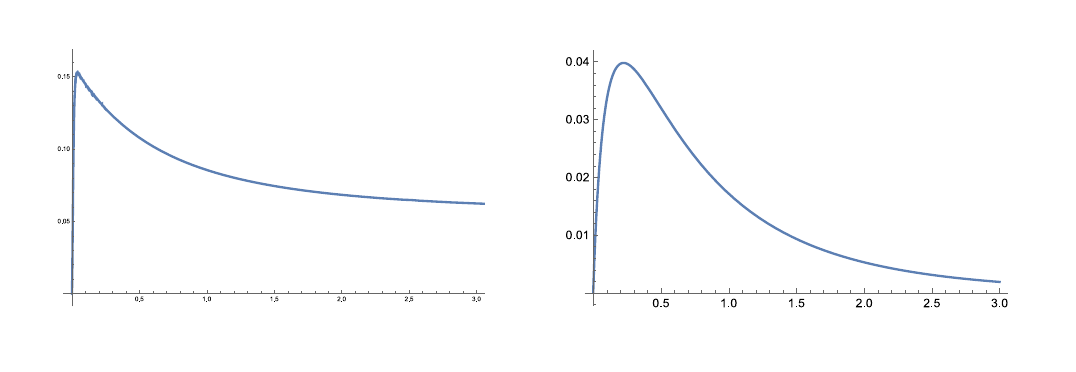}}%
    \put(0.24779392,0.02866034){\color[rgb]{0.0745098,0.03529412,0.03529412}\makebox(0,0)[lt]{\lineheight{1.25}\smash{\begin{tabular}[t]{l}$\text{(a)}$\end{tabular}}}}%
    \put(0.7364908,0.02708886){\color[rgb]{0.0745098,0.03529412,0.03529412}\makebox(0,0)[lt]{\lineheight{1.25}\smash{\begin{tabular}[t]{l}$\text{(b)}$\end{tabular}}}}%
    \put(0.94845351,0.07085728){\color[rgb]{0.15294118,0.06666667,0.06666667}\makebox(0,0)[lt]{\lineheight{1.25}\smash{\begin{tabular}[t]{l}$m^2$\end{tabular}}}}%
    \put(0.55969033,0.30254514){\color[rgb]{0.15294118,0.06666667,0.06666667}\makebox(0,0)[lt]{\lineheight{1.25}\smash{\begin{tabular}[t]{l}$\delta m^2_{\sigma}$\end{tabular}}}}%
    \put(0.0759145,0.30254514){\color[rgb]{0.15294118,0.06666667,0.06666667}\makebox(0,0)[lt]{\lineheight{1.25}\smash{\begin{tabular}[t]{l}$\delta m^2_{\sigma}$\end{tabular}}}}%
    \put(0.46172782,0.07184056){\color[rgb]{0.15294118,0.06666667,0.06666667}\makebox(0,0)[lt]{\lineheight{1.25}\smash{\begin{tabular}[t]{l}$m^2$\end{tabular}}}}%
  \end{picture}%
\endgroup%

    \caption{The dependence of $\delta m^2_{\sigma}$ on the squared mass of the matter fields. The first plot (a) depicts the numerical value for scalar field and the second plot (b) is analytical approximation for fermions.}
    \label{fig:2d_fermions}
\end{figure}

Consider the standard kinetic term for Dirac fermions in two dimensions (we will follow the article \cite{Stahl:2015gaa}):
\beq\label{eq:ferm_in_ds}
    S_{\text{ferm}} = \int d^2x \sqrt{|g|}\frac{i}{2}\left[\Bar{\psi}\underline{\gamma}^{\mu}\nabla_{\mu}\psi - \nabla^{\mu}\Bar{\psi}\underline{\gamma}^{\mu}\psi - m \Bar{\psi}\psi\right],
\eeq
where $\underline{\gamma}^{\mu}$ are gamma-matrices in curved space, and the action of covariant derivatives are determined by spin-connection, see \cite{Stahl:2015gaa} for details. In flat space the gamma matrices are chosen in the form 
\beq\label{eq:gamma_matrices}
    \gamma^{0} = \begin{pmatrix}
        0 & 1\\
        1 & 0
    \end{pmatrix};\;\;
    \gamma^{1} = \begin{pmatrix}
        0 & 1\\
        -1 & 0
    \end{pmatrix}.
\eeq
Then we quantize the fermionic field with canonical anticommutation relation conditions:
\beq \label{eq:fermions_decomposition}
    \begin{aligned}
        \psi(t,\mathbf{x})=\int\dfrac{d p}{2\pi}e^{ipx}\left[
\widehat{b}_{p}\psi^{(+)}_{p}(t)+
\widehat{d}^{\dagger}_{-p}\psi^{(-)}_{p}(t)\right], \quad 
\left\{\widehat{b}_{p},\widehat{b}_{q}^\dagger 
\right\} = \left\{\widehat{d}_{p},\widehat{d}_{q}^\dagger 
\right\} = 2\pi \delta(p-q),
\end{aligned}
\eeq
where we denote by $\psi^{(+)}_{p}(t),\;\psi^{(-)}_{p}(t)$ the positive- and negative-frequency solutions of Dirac equation, determined by (\ref{eq:ferm_in_ds}). Then, using the Schwinger-Keldysh technique for fermions \cite{Akhmedov:2019rvx, kamenev_2011}, we obtain for the effective mass of $\sigma$:
\beq\label{eq:ferm_sigma_mass}
    m_{\sigma}^2 = 8m^2\int_{-\infty}^{\infty}dt'\sqrt{|g(t')|}\int_{-\infty}^{\infty}\frac{dp}{2\pi}\text{Im}\left\{\Bar{\psi}^{(+)}\left(p|t\right)\psi^{(-)}\left(p|t\right)\Bar{\psi}^{(-)}\left(p|t'\right)\psi^{(+)}\left(p|t'\right)  \right\}.
\eeq
In the case of flat space $\psi^{(+)}\left(p|t\right) = e^{-i\sqrt{p^2+M^2}t}\begin{pmatrix}
        0\\
        1
    \end{pmatrix}, \; \psi^{(-)}\left(p|t\right) = e^{i\sqrt{p^2+M^2}t}\begin{pmatrix}
        1\\
        0
    \end{pmatrix}$ we encounter logarithmic divergence:
\beq\label{eq:ferm_Flat_sigma_mass}
    m_{\sigma}^2 = \frac{4m^2}{\pi}\int_{0}^{\infty}\frac{dp}{\sqrt{p^2+M^2}},
\eeq
which can be connected to the additional non-local terms, which appear in deformation of Mabuchi action for fermions \cite{Bilal:2021fqq}. Nevertheless, as we are interested in the difference $\delta m^2_{\sigma} \equiv m^2_{\sigma}\bigg|_{\text{dS}} - m^2_{\sigma}\bigg|_{\text{Mink}}$, let us proceed naively and just subtract the UV region in (\ref{eq:ferm_sigma_mass}). As the UV behaviour of the harmonics is the same independently of the curvature, one will have the same logarithmic divergence. In dS the mode functions are as follows:
\beq
\begin{aligned}
    \psi^{(+)}\left(k|\eta\right) = \sqrt{\frac{H}{2|k|}}
    \begin{cases}
        \begin{pmatrix}
        -im W_{-\frac{1}{2},im}\left(2i|k|\eta\right),\\
        W_{\frac{1}{2},im}\left(2i|k|\eta\right)
    \end{pmatrix}, k>0
    \\
        \begin{pmatrix}
        W_{\frac{1}{2},im}\left(2i|k|\eta\right),\\
        -im W_{-\frac{1}{2},im}\left(2i|k|\eta\right)
    \end{pmatrix}, k<0
    \end{cases}
\\
    \psi^{(-)}\left(k|\eta\right) = \sqrt{\frac{H}{2|k|}}
    \begin{cases}
        \begin{pmatrix}
        W_{\frac{1}{2},im}^*\left(2i|k|\eta\right),\\
        -im W_{-\frac{1}{2},im}^*\left(2i|k|\eta\right)
    \end{pmatrix}, k>0
    \\
        \begin{pmatrix}
        -im W_{-\frac{1}{2},im}^*\left(2i|k|\eta\right),\\
        -W_{\frac{1}{2},im}^*\left(2i|k|\eta\right)
    \end{pmatrix}, k<0
    \end{cases}
\end{aligned}
\eeq
where $W_{\kappa,\mu}(z)$ is a Whittaker function. Hence, after the subtraction of the UV region in (\ref{eq:ferm_sigma_mass}) the only contributing range of integration over physical momentum is $\frac{1}{2\tau}<|k|\eta<\frac{1}{2},\;\tau = \frac{\eta'}{\eta}$, so that with the use of asymptotics of Whittaker function we can estimate:
\beq\label{eq:delta_m_sigma_ferm}
\begin{aligned}
    &\delta m^2_{\sigma} \simeq 
    \\
    &\simeq \frac{32m^2}{\pi}\int_{1}^{\infty}\frac{d\tau}{\tau}\int_{\frac{1}{2\tau}}^{\frac{1}{2}}\text{Im}\left(\left[\text{Re}^2\left\{\frac{\Gamma(2im)}{\Gamma(im)}\frac{1}{\left(2i\xi\right)^{im}}\right\} + m^2\text{Re}^2\left\{\frac{\Gamma(2im)}{im\Gamma(im)}\frac{1}{\left(2i\xi\right)^{im}}\right\} \right]e^{-2i\xi\tau} \right). 
\end{aligned}
\eeq
The integral (\ref{eq:delta_m_sigma_ferm}) can be calculated and we find
\beq\label{eq:sigma_mass_ferm_finally}
    \delta m^2_{\sigma} \simeq 4\left(1-\frac{2\text{Si}(1)}{\pi}\right)\frac{m^2}{4^m}\frac{\sinh{\left(\pi m\right)}}{\sinh{\left(2\pi m\right)}},
\eeq
where $\text{Si}(x)$ is the sine integral function. The qualitative dependence of $\delta m^2_{\sigma}$ on the mass of the matter field is the same as for the scalar field, including the sign of this quantity (see fig.\ref{fig:2d_fermions}), which confirms our discussion above. Note also that this answer is determined by the IR behaviour of the mode functions. 

As we already noted, we define the mass term for $\sigma$ in the effective action with the wrong sign, so that the answers depicted on figures \ref{fig:2d_Plot} and \ref{fig:2d_fermions} describe the positive mass in the Klein-Gordon equation, because the central charge for the scalar and fermion matter is $1$ and $\frac{1}{2}$ correspondingly:
\beq\label{eq:KG_2D}
    \Box\sigma + \frac{96\pi}{c}\delta m^2_{\sigma}\sigma + \ldots = 0.
\eeq
Let us stress here again that despite the expansion over small momenta, the ellipsis here account for non-perturbative contributions such as the Mabuchi action (\ref{eq:Mabuchi_action}), whose significance at different energy scales is yet to be understood in more detail. Our conjecture is that $\delta m^2_{\sigma} = m^2_{\sigma}\bigg|_{\text{dS}} - m^2_{\sigma}\bigg|_{\text{Mink}}$ describes the response of quantum matter to the curved background at low energy scale. Further, if we treat gravity at quantum level in  two dimensions, we should change $c\rightarrow c-26$ in (\ref{eq:KG_2D}), including into account the central charge of ghosts \cite{Polyakov:1981rd}. In this case the kinetic term in the effective action can have the correct sign, and the $\delta m_{\sigma}^2$ corresponds to the negative squared mass in the KG equation (\ref{eq:KG_2D}), effectively describing a distortion of the initial background. However, the more accurate understanding of the evolution of metric perturbations requires further study of the effective action. 
\section{Conclusion}\label{sec:Conclusion}
The paper discusses the concept of massive terms in the effective gravitational action, which are induced by quantum fields of matter in dS background. Although it is known that the initial state of quantum field theory in dS must decay due to particle production \cite{Akhmedov:2019esv, Polyakov:2022dpa}: $\big\langle\text{Out}\big|\text{In}\big\rangle \neq 1$, the comprehensive understanding of the physical consequences of this phenomena requires the consideration of the response of specific systems to external influences. Furthermore, the evolution of metric perturbations in the presence of classical stress-energy tensor can tell us a lot about the physics in the early Universe \cite{Gorbunov:1354521, gorbunov2011introduction}, while it is also well understood in physical community that loop corrections in dS can lead to drastic modifications of tree-level results \cite{Akhmedov:2017ooy, Akhmedov:2019cfd,Akhmedov:2012pa, Akhmedov:2014doa, Akhmedov:2013vka}.

It turns out that spin-$2$ part of metric perturbations, associated with the graviton, does not acquire a mass in one-loop effective action. Although the notion of mass in an expanding space-time enables us to work in non-equilibrium framework and include some counterterms to the effective action, we argue that this result is regularisation scheme-independent. This is due to the auxiliary argument, that there is no ``exchange'' of the Goldstone vector, which gives mass to the graviton, because it cannot be produced by the stress-energy tensor of free field theory out of the initial Bunch-Davies vacuum. It would be very interesting to generalize our observations to self-interacting theories, where we must take into account loop corrections \cite{Akhmedov:2013vka}. Additionally, on physical grounds, we must investigate a much larger scope of various initial states. Indeed, Bunch-Davies vacuum preserves the highest number of symmetries of the problem, while in any real situation, most symmetries are broken. For instance, in the context of the early Universe, the most natural initial state is thermal state with non-canonical temperature \cite{Diakonov:2023hzg, Bazarov:2021rrb, Alonso-Monsalve:2023jfq}. In this case, the meanvalue of stress-energy tensor isn't proportional to the metric $\big\langle :T_{\mu\nu}:\big\rangle \neq \delta\lambda g_{\mu\nu} $ \cite{Diakonov:2023hzg, Bazarov:2021rrb}, hence one should expect additional contributions from tadpole diagrams on the fig.\ref{fig:Effective_action}. Furthermore, in global dS the isometry group is broken due to divergent IR behaviour of the propagators at past infinity \cite{Krotov:2010ma, Akhmedov:2012dn}, making one to introduce a Cauchy surface at finite time. Our analysis in sections \ref{sec:Definitions}--\ref{sec:Tensor_mode} can be extended to more general geometries as FLRW and global dS, so we also leave such questions for further investigations.

Finally, in Section \ref{sec:Scalar_sector} we have outlined the problems we encounter when considering the scalar sector of gravitational perturbations at the one-loop level. The question of “scalar ghost” in gravity remains unresolved \cite{Zakharov:1970cc, deRham:2014zqa, Hinterbichler:2011tt}, but the consideration of loop modified gravity at classical level will allow us to improve predictions about the behavior of matter in the Universe at early stages incorporating the evolution of scalar metric perturbations. In the simplest case of two dimensions, we observe that light matter fields exhibit a considerable response to an expanding background, because these fields also have significantly different infrared behavior. Therefore, we believe that further study of this issue using various approaches will provide answers about the stability and behavior of matter in expanding gravitational backgrounds.
\section*{Acknowledgments}
We would like to acknowledge discussions with D. V. Diakonov, K. V. Bazarov and D. A. Trunin. We are grateful to Emil Akhmedov for valuable discussions, careful reading of the paper, correcting the text and support. Especially we thank Fedor K. Popov for initiating this work, fruitful ideas and careful reading of the paper. This work was supported by the grant from the Foundation for the Advancement of Theoretical Physics and Mathematics ``BASIS''.
\appendix
\section{Bubble diagram}\label{subSec:Bubble}
In this subsection we analyze the expression for the non-local part $\Pi_{\mu\nu|\alpha\beta}^{\text{bub}}$ of the bubble diagram. First of all, let us implement the $O(D-1)$ rotational symmetry of dS, rotational symmetry of the chosen quantum state and the property $\Pi_{\mu\nu|\alpha\beta}^{\text{bub}}\left(\mathbf{k}|\eta, \eta'\right) = -\Pi_{\alpha\beta| \mu\nu}^{\text{bub}}\left(-\mathbf{k}|\eta', \eta\right)$ to write the general expression for the polarization operator:
\beq \label{eq:rotationally_inv_decomposition}
\begin{aligned}
    \Pi_{00|00}^{\text{bub}} &= a, \;\; \Pi_{00|0k}^{\text{bub}} = i\frac{k_{k}}{k^2}b,
    \\
    \Pi_{00|kl}^{\text{bub}} &= f'_{1}\delta^{\bot}_{kl} + f'_{2}\frac{k_{k}k_{l}}{k^4} = f_{1}\delta^{\bot}_{kl} + f_2\frac{k_{k}k_{l}}{k^4} + f\left( \delta_{kl} - (D-1)\frac{k_{k}k_{l}}{k^2} \right),
    \\
    \Pi_{0i|0k}^{\text{bub}} &= c_{1} \delta^{\bot}_{ik} + c_{2}\frac{k_{i}k_{k}}{k^4},
    \\
    \Pi_{0i|kl}^{\text{bub}} &= i\frac{k_{i}}{k^2}d'_{1}\delta^{\bot}_{kl} + id'_2\frac{k_{i}k_{k}k_{l}}{k^6} + id_{3}\left[\delta^{\bot}_{il}\frac{k_k}{k^2} + \delta^{\bot}_{ik}\frac{k_l}{k^2}  \right] \equiv
    \\
    &\equiv i\frac{k_{i}}{k^2}d_{1}\delta^{\bot}_{kl} + id_{2}\frac{k_{i}k_{k}k_{l}}{k^6} + id\left( \delta_{kl} - (D-1)\frac{k_{k}k_{l}}{k^2} \right) + id_{3}\left[\delta^{\bot}_{il}\frac{k_k}{k^2} + \delta^{\bot}_{ik}\frac{k_l}{k^2}  \right],
    \\
    \Pi_{ij|kl}^{\text{bub}} &= e_{1}\delta^{\bot}_{ij}\delta^{\bot}_{kl} + \left[ -\overline{e}_2\frac{k_{i}k_{j}}{k^4}\delta^{\bot}_{kl} + e_{2}\frac{k_{k}k_{l}}{k^4}\delta^{\bot}_{ij} \right] + e_{3}\frac{k_{i}k_{j}k_{k}k_{l}}{k^8} + 
    \\
    &+ e_{4}\left[ \delta^{\bot}_{ik}\frac{k_{j}k_{l}}{k^4} + \delta^{\bot}_{jk}\frac{k_{i}k_{l}}{k^4} + \delta^{\bot}_{il}\frac{k_{j}k_{k}}{k^4} + \delta^{\bot}_{jl}\frac{k_{i}k_{k}}{k^4} \right] + e_{5}\left[ \delta^{\bot}_{ik}\delta^{\bot}_{jl} + \delta^{\bot}_{il}\delta^{\bot}_{jk} \right],
\end{aligned}
\eeq
where all the coefficient functions $a, b, f\text{'s}, c\text{'s}, d\text{'s}, e\text{'s}$ depend on $\eta,\eta',\mathbf{k}$ and we have also used line to denote switching variables, e.g., $\overline{a}(\mathbf{k}|\eta,\eta') \overset{\text{def}}{=} a(\mathbf{k}|\eta',\eta)$; $\delta^{\bot}_{kl} \overset{\text{def}}{=} \delta_{kl} - \frac{k_{k}k_{l}}{k^2}$. In (\ref{eq:rotationally_inv_decomposition}) we encounter 14 coefficients, but they are not independent because of the Ward identities $\hat{\nabla}_{\alpha}\Pi_{\mu\nu|\alpha\beta}^{\text{bub}} = 0$, which can be written explicitly as follows:
\begin{equation}\label{eq:Ward_identities}
        \nabla_{\eta'}\Pi_{\mu\nu|00}^{\text{bub}} - (-ik_{k})\Pi_{\mu\nu|0k}^{\text{bub}} + \frac{1}{\eta'}\left(\Pi_{\mu\nu|00}^{\text{bub}} - \Pi_{\mu\nu|kk}^{\text{bub}}\right) = 0, \quad
        \nabla_{\eta'}\Pi_{\mu\nu|0k}^{\text{bub}} - (-ik_{l})\Pi_{\mu\nu|kl}^{\text{bub}} = 0.
\end{equation}
The solution of (\ref{eq:Ward_identities}) can be chosen in the following form:
\begin{equation}\label{eq:Solution_of_Ward_identites}
\begin{aligned}
     a &= a_1 - a_2 + \frac{m^2}{\eta'^2}a_3, \quad b = \nabla_{\eta'}a - \frac{D-2}{\eta'}\left(a_1 + a_2 - \frac{m^2}{\eta'^2}a_3 \right) + \frac{2}{\eta'}\frac{m^2}{\eta'^2}a_{3}, 
    \\
    c_2 &= -\nabla_{\eta'}\overline{b} +\frac{D-2}{\eta'}\overline{b} + \frac{D-1}{\eta'}b_1 + \frac{D-2}{\eta'}b_2, 
    \\
    f_1 &= a_1 + a_2 - \frac{m^2}{\eta'^2}a_3, \quad f_2 = k^2\left[a_1 - a_2 - \frac{m^2}{\eta'^2}a_3\right], 
    \\
    f &= \frac{1}{k^2}\frac{1}{D-2}\left(f_2 + \frac{1}{k^2}\nabla_{\eta'}b \right) \;\; \left(\text{hence } f'_1 = f_1 + f, \; f'_2 = -\nabla_{\eta'}b  \right),
    \\
    d_1 &= \overline{b} + b_1 + b_2, \quad d_2 = k^2\left( \overline{b} + b_1 \right), \; d_3 = \nabla_{\eta'}c_1, 
    \\
    d &= \frac{1}{k^2}\frac{1}{D-2}\left( d_2 - \nabla_{\eta'}c_2 \right) \;\; \left(\text{hence } d'_1 = d_1 + d, \; d'_2 = \nabla_{\eta'}c_2 \right),
    \\
    e_2 &= -\nabla_{\eta'}\overline{d}_1, \quad e_3 = -\nabla_{\eta'}\overline{d}_2, \quad e_4 = -\nabla_{\eta'}\overline{d}_3,
    \\
    e_1 &= -\frac{2}{D-2}e_5 - \frac{1}{k^2}\frac{1}{D-2}e_2 - \frac{1}{D-2}\overline{f'_1} - \eta'\left(\nabla_{\eta'}\overline{f'_1} + \overline{d'_1} \right).
\end{aligned}
\end{equation}
Then we are left with 7 independent coefficient functions $a_1, a_2, a_3, b_1, b_2, c_1, e_5$. The direct calculation of the stress-energy correlators (\ref{eq:Bubble_contribution}) gives the following expressions:
\begin{equation}\label{eq:7_independent_coefficients}
\begin{aligned}
    a_1 &= \frac{1}{2}\int_{p} \left[ \partial_{\eta}\partial_{\eta'}F\partial_{\eta}\partial_{\eta'}\rho - \left(\mathbf{kp} - p^2 -\frac{m^2}{\eta^2}\right)\partial_{\eta'}F\partial_{\eta'}\rho \right],
    \\
    a_2 &= \frac{1}{2}\int_{p}\left(\mathbf{kp} - p^2\right)\left[ \partial_{\eta}F\partial_{\eta}\rho - \left(\mathbf{kp} - p^2 -\frac{m^2}{\eta^2}\right)F\rho \right],
    \\
    a_3 &= \frac{1}{2}\int_{p}\left[ \partial_{\eta}F\partial_{\eta}\rho - \left(\mathbf{kp} - p^2 -\frac{m^2}{\eta^2}\right)F\rho \right],
    \\
    b_1 &= -2\frac{m^2}{\eta'^2}\times\frac{1}{2}\int_{p}\bigg[(k^2-\mathbf{kp})\partial_{\eta}F\rho + \mathbf{kp} F\partial_{\eta}\rho\bigg],
    \\
    b_2 &= 2\times\frac{1}{2}\int_{p} \left(\mathbf{kp} - p^2\right) \bigg[(k^2-\mathbf{kp})\partial_{\eta}F\rho + \mathbf{kp} F\partial_{\eta}\rho \bigg],
    \\
    c_1 &= \frac{1}{D-2}\times\frac{1}{2}\int_{p}\left(p^2 -\frac{(\mathbf{kp})^2}{k^2}\right)\bigg[\partial_{\eta}\partial_{\eta'}F\rho + F\partial_{\eta}\partial_{\eta'}\rho - \partial_{\eta'}F\partial_{\eta}\rho - \partial_{\eta}F\partial_{\eta'}\rho \bigg],
    \\
    e_5 &= \frac{4}{D(D-2)}\times\frac{1}{2}\int_{p} \left(p^2 -\frac{(\mathbf{kp})^2}{k^2}\right)^2 F\rho,
\end{aligned}
\end{equation}
where for brevity we've denoted $F = F(\mathbf{p}|\eta,\eta'),\;\rho=\rho(\mathbf{k}-\mathbf{p}|\eta,\eta')$ and $\int_{p} = \int\frac{d^{D-1}\mathbf{p}}{(2\pi)^{D-1}}$. Thus, equations (\ref{eq:rotationally_inv_decomposition}), (\ref{eq:Solution_of_Ward_identites}), (\ref{eq:7_independent_coefficients}) give the whole contribution of the non-local part to the bubble diagram through several loop integrals.
\section{Gauge invariance of the effective action}\label{App:Gauge_inv}
Let us pick up only the contributions to $\delta_{\xi}\Gamma^{\text{eff}}$, which contain $h_{q}^{00}$ in the order $\mathcal{O}\left(||h_{q}\cdot\xi||\right)$. For the ``source''-term in (\ref{eq:effective_action}) we take the contributions with $h^{\alpha\beta}$ in (\ref{eq:gauge_transformation}) and get:
\beq
\begin{aligned}
    -\frac{1}{2}\delta_{\xi}\left( h_{q}^{\alpha\beta}\left\langle T_{\alpha\beta}^{cl-cl} \right\rangle \right ) &= 
    \\
    = h_{q}^{00}\partial_{\eta}\xi^0&\left\langle T_{00}^{cl-cl} \right\rangle - \frac{2}{\eta}h_{q}^{00} \xi^0\left\langle T_{00}^{cl-cl} \right\rangle + \frac{1}{2}\partial_{\eta}h_{q}^{00} \xi^0\left\langle T_{00}^{cl-cl} \right\rangle  + \frac{1}{2}\partial_{i}h_{q}^{00} \xi^i\left\langle T_{00}^{cl-cl} \right\rangle,
\label{A1}
\end{aligned}
\eeq
where we also use that in the rotationally invariant state one has $ \left\langle T_{0i}^{cl-cl} \right\rangle = 0,\; \partial_{i}F(\underline{x},\underline{x}) = 0$, etc. The variation of the local parts of (\ref{eq:effective_action}) in the required order is
\begin{gather}
    \delta_{\xi}\left( h_{cl}^{\mu\nu}(x)\Pi^{\text{loc}}_{\mu\nu|\alpha\beta}(\eta)h_{q}^{\alpha\beta}(x) \right ) = -\frac{1}{2}\left[ \left( \partial_{\eta}\xi^{0} - \frac{1}{\eta}\xi^{0} \right)\left\langle T_{00}^{cl-cl} \right\rangle + \left( \partial_{i}\xi^{j} - \frac{1}{\eta}\delta^{j}_{i}\xi^{0} \right)\left\langle T_{ij}^{cl-cl} \right\rangle \right]h_{q}^{00}.
\label{A2}
\end{gather}
Finally, we must take into account terms with the derivatives of the theta-functions in (\ref{eq:effective_action}) after the gauge variation and integration by parts, while the derivatives of the polarization operator in this term is vanishing due to the Ward identities (\ref{eq:Ward_identities}). Again, we use the commutation relations (\ref{eq:canonical_commutator}) to find the bubble contribution at coincident points and obtain in the order under consideration:
\begin{gather}
    -\frac{1}{2}\delta_{\xi}\int h_{q}\Pi^{\text{bub}}h_{cl} = \frac{1}{2}\int\frac{d^Dx}{H^D\eta^D}h_{q}^{00}\bigg[ \partial_{i}\xi^{i}\partial_{\eta}\partial_{\eta'}F(x,x) + \partial_{i}\xi^{j}\partial_{x^i}\partial_{y^j}F(x,x) \bigg],
\label{A3}
\end{gather}
(where the derivatives over $\eta'$ and $y^j$ are referred to the second argument of the Keldysh function). Therefore, the whole variation in this order vanishes:
\begin{equation}\label{A4}
    \delta_{\xi}\Gamma_{\text{eff}} = -\frac{1}{2}\int\frac{d^Dx}{H^D\eta^D}h_{q}^{00}(x)\xi^{0}(x)\hat{\nabla}_{\mu}\left\langle T_{\mu0}^{cl-cl} \right\rangle = 0.
\end{equation}
In the similar way one can check that all other components of $\delta_{\xi}\Gamma_{\text{eff}}$ vanish.
\section{Integral relation for Green functions}\label{App:Keldysh_function}
Consider the bare action for the scalar field (\ref{eq:bare_action}) and the interaction term in the form:
\begin{equation}\label{B1}
    \delta_{p^2}S_{\text{int}} = -\int_{p}\int\frac{d\eta}{H^{D-2}\eta^{D-2}}\delta p^2\phi_{cl}(\mathbf{p},\eta)\phi_{q}(-\mathbf{p},\eta) = -\delta p^2\int d^{D}x\sqrt{\hat{g}}\phi_{cl}(x)\phi_{q}(x).
\end{equation}
Then the first-order correction to the exact Keldysh propagator is given by
\begin{gather}
    \delta_{p^2}F(\mathbf{p}|\eta,\eta) = -i\delta p^2\int d^{D-1}\mathbf{x}d^{D-1}\mathbf{z}\frac{d\eta'}{H^{D-2}\eta'^{D-2}}\left\langle \phi_{cl}(\mathbf{x},\eta) \phi_{cl}(z)\phi_{q}(z) \phi_{cl}(\mathbf{y},\eta)\right\rangle e^{-i\mathbf{p}(\mathbf{x}-\mathbf{y})} = 
    \nonumber
    \\
    = -\delta p^2 2\int_{\eta}^{\infty} \frac{d\eta'}{H^{D-2}\eta'^{D-2}}\rho(\mathbf{p}|\eta,\eta')F(\mathbf{p}|\eta,\eta').
\label{B2}    
\end{gather}
On the other hand, we have $\delta_{p^2}F(\mathbf{p}|\eta,\eta) = \delta p^2\partial_{p^2}F(\mathbf{p}|\eta,\eta)$ by construction and the relation (\ref{eq:relation_for_Keldysh}) follows immediately.
\section{Relations for the stress-energy tensor}\label{App:Relations_for_STE}
From the very definition (\ref{eq:Stress-energy_tensor}) we have the relation
\beq\label{eq:Trace_stress_tensor}
    T_{ii} = -\left(D-1\right)T_{00} + \left(D-1\right)\partial_{\eta}\phi\partial_{\eta}\phi + \partial_{i}\phi\partial_{i}\phi \equiv -\left(D-1\right)T_{00} + \Delta T.
\eeq
In addition, the covariant conservation condition $\hat{\nabla}^{\mu}T_{\mu\nu} = 0$ reads
\beq\label{eq:conserv_stress_tensor}
    \hat{\nabla}^{\mu}T_{\mu\nu} = \partial_{\eta}T_{00} + \frac{2}{\eta}T_{00} + \partial^{i}T_{i0} - \frac{1}{\eta}\Delta T = 0.
\eeq
In momentum space in the limit $k\rightarrow 0$ the last equation gives
\beq\label{eq:DeltaT_through_T00}
    \Delta T\left(\mathbf{k} = \mathbf{0},\eta\right) = \left[\eta\partial_{\eta} + 2\right]T_{00}\left(\mathbf{0}|\eta\right).
\eeq
In $\mathbf{x}$-space language the limit $k\rightarrow 0$ is equivalent to the integral over $\mathbf{x}$. Then first of all we can derive the equality (below we write the stress-energy tensor in position space and use translational symmetry of the correlator):
\beq\label{eq:1relation_for_STE}
    \begin{aligned}
        \int_{\eta}^{\infty}\frac{d\eta'}{\eta'^{D-2}}\left\langle BD\bigg|\left[T_{00}\left(\mathbf{0}, \eta\right), \int_{\mathbf{x}}\Delta T\left(\mathbf{x}, \eta'\right)\right] \bigg|BD\right\rangle = 
        \\
        =\int_{\eta}^{\infty}\frac{d\eta'}{\eta'^{D-2}}\left(\eta'\partial_{\eta'} + 2  \right)\left\langle BD\bigg|\left[T_{00}\left(\mathbf{0}, \eta\right), \int_{\mathbf{x}} T_{00}\left(\mathbf{x}, \eta'\right)\right] \bigg|BD\right\rangle = 
        \\
        = \left(D-1\right)\int_{\eta}^{\infty}\frac{d\eta'}{\eta'^{D-2}}\left\langle BD\bigg|\left[T_{00}\left(\mathbf{0}, \eta\right), \int_{\mathbf{x}} T_{00}\left(\mathbf{x}, \eta'\right)\right] \bigg|BD\right\rangle =
        \\
        =\left(D-1\right)\int_{\eta}^{\infty}\frac{d\eta'}{\eta'^{D-2}}\Pi_{00|00}\left(\mathbf{0}|\eta,\eta'\right),
    \end{aligned}
\eeq
where in the last line we have integrated by parts. We also can use the invariance of the correlation functions in $\mathbf{x}$-space under transformations $\mathbf{x}\rightarrow a\mathbf{x},\;\eta\rightarrow a\eta$ (indeed, the correlators depend on the dS-invariant variable $Z = \frac{\eta^2 + \eta'^2 - \left(\mathbf{x}-\mathbf{y}\right)^2}{2\eta\eta'}$ \cite{Akhmedov:2019esv, Allen:1985wd}) and introduce new variables $y = ax, \; \omega = \frac{\eta^2}{\eta'}$ to obtain:
\beq\label{eq:2relation_for_STE}
    \begin{aligned}
        \int_{\eta}^{\infty}\frac{d\eta'}{\eta'^{D-2}}\left\langle BD\bigg|\left[\int_{\mathbf{x}}\Delta T\left(\mathbf{x}, \eta\right), T_{00}\left(\mathbf{0}, \eta'\right)\right] \bigg|BD\right\rangle = 
        \\
        =\int_{\eta}^{\infty}\frac{d\eta'}{\eta'^{D-2}}\frac{a^4}{a^{D-1}}\left\langle BD\bigg|\left[\int_{\mathbf{y}}\Delta T\left(\mathbf{y}, a\eta\right), T_{00}\left(\mathbf{0}, a\eta'\right)\right] \bigg|BD\right\rangle = 
        \\
        = \frac{1}{\eta^D}\int_{0}^{\eta}\omega d\omega \left\langle BD\bigg|\left[\int_{\mathbf{y}}\Delta T\left(\mathbf{y}, \omega\right), T_{00}\left(\mathbf{0}, \eta\right)\right] \bigg|BD\right\rangle =
        \\
        =\frac{1}{\eta^D}\int_{0}^{\eta}\omega d\omega\left(\omega\partial_{\omega} + 2\right) \left\langle BD\bigg|\left[\int_{\mathbf{y}} T_{00}\left(\mathbf{y}, \omega\right), T_{00}\left(\mathbf{0}, \eta\right)\right] \bigg|BD\right\rangle \equiv 0,
    \end{aligned}
\eeq
where we again integrate by parts in the last line and in the second lime we take $a = \frac{\eta}{\eta'}$. With the use of the derived equations we find
\beq
    \int_{\eta}^{\infty}\frac{d\eta'}{\eta'^{D-2}}\Pi_{00|ii}\left(\mathbf{0}|\eta,\eta'\right) = 0,\; \int_{\eta}^{\infty}\frac{d\eta'}{\eta'^{D-2}}\Pi_{ii|00}\left(\mathbf{0}|\eta,\eta'\right) = -\left(D-1\right)\int_{\eta}^{\infty}\frac{d\eta'}{\eta'^{D-2}}\Pi_{00|00}.
\eeq
\section{The mass of photon in \texorpdfstring{$\text{AdS}_4$}{Lg}}\label{App:Photon_AdS}
 It is convenient for us to treat $\text{AdS}_4$ from the beginning as a hyperboloid, embedded into the five-dimensional pseudo-Euclidean space with coordinates $X^A$: $\eta_{AB}X^AX^B = L^2$, where $\eta_{AB} = \text{diag}\left(1,1,-1,-1,-1\right)$. Following the approach developed in the papers \cite{Janssen:1986fz, Akhmedov:2020jsi}, we first write the bare action in the form:
 \begin{gather}
    S[B^A,\phi] = \int d\mu_{X}\left[-\frac{1}{4}F_{AB}F^{AB} + \left| \left(\partial^{0}_A + ieB_{A} \right)\phi \right|^2 -m^2\left|\phi\right|^2\right],
\label{bare_action_QED_AdS}
 \end{gather}
 where $d\mu_{X} = 2L\delta(X^2-L^2)d^5X$ is the AdS-invariant measure, $\partial^{0}_{A} = \partial_{A} - \frac{1}{X^2}X_{A}X^{I}\partial_{I}$ is the tangent derivative, $F_{AB} = \left( \partial^{0}_{A} - X_{A} \right)B_{B} - \left( \partial^{0}_{B} - X_{B} \right)B_{A}$ and $B^{A}$ is a vector potential, which is considered to be tangent to the hyperboloid: $X^{A}B_{A} = 0,\;X^{A}\in \text{AdS}_4$. The vector potential in the $\text{AdS}_4$ are obtained by a pull--back of $B_A$. Let us impose additional transversal condition $\partial_{A}B^{A}_{t} = 0$, so that the free equation of motion for the vector--potential simplifies to the ordinary wave--equation:
 \begin{equation}\label{free_EOM_for_B}
     \left[X^2\partial^2 - \left(X\cdot \partial\right)^2 - 3\left(X\cdot \partial\right) - 2\right]B^{A}_{t} = 0.
 \end{equation}
 This is the wave-equation for the fields in the massless representation $D(2,1)$ \cite{Fronsdal:1978vb, Nicolai:1984hb}. The corresponding gauge variation \cite{Fronsdal:1978vb} $\delta_{\varphi}B_{A} = X^2\partial_{A}\varphi - X_{A}X^{I}\partial_{I}\varphi$ is determined by the scalar $\varphi\in D(3,0)$. We can construct the projector onto the transversal vector-potential $B^{A}_{t}$ by the following gauge transformation:
 \begin{gather}
    B^{A}_{t} = \hat{\mathscr{P}}^{AB}_{t}B_{B} = B^{A} - \left[ X^2\partial^{A}\varphi - X^{A}X^{I}\partial_{I}\varphi  \right],
    \nonumber
    \\
    \varphi = \frac{1}{ X^2\partial^2 - \left(X\cdot \partial\right)^2 - 3\left(X\cdot \partial\right)}\partial^{C}B_{C}.
\label{projector_on_Bt}
 \end{gather}
 Connecting the ordinary mass-term in (\ref{bare_action_QED_AdS}) with the parameter $E$ in four-dimensional case $m^2L^2 = E(3-E)$ we write the equation for the Wightman function:
 \begin{equation}\label{equation_wightman_function_AdS}
     \left[\left(1-Z^2\right)\partial^2_{Z} - 4Z\partial_{Z} + E(3-E)\right]W_{E}(Z) = 0,
 \end{equation}
 where $Z(X,Y) = \frac{X^MY_M}{L^2}$ is invariant variable. The solution for $E\neq 1,\;2$ is \cite{Porrati:2001db, Allen:1985wd}:
 \begin{equation}\label{wightman_AdS}
     W_{E}(Z) = \frac{1}{4\pi^2L^2}\frac{\Gamma(E)\Gamma(E-1)}{\Gamma(2E-2)}\frac{1}{Z^E}\;_1F_2\left(E,E-1; 2E-2;\frac{1}{Z}\right).
 \end{equation}
 For conformally coupled scalar $E = 1,\;2$ the solution looks like
 \begin{equation}\label{conf_wightman_AdS}
     W_c(Z) = \frac{1}{4\pi^2L^2}\left(\alpha\frac{1}{Z^2-1} + \beta\frac{Z}{Z^2-1}\right),
 \end{equation}
 where the choice of $\alpha,\;\beta$ corresponds to a different boundary conditions. The Feynman propagator can be obtained by the introduction of $i\epsilon$-prescription: $G_{F}(X,Y) = W(Z+i\epsilon)$. The kernel of the inverse operator in (\ref{projector_on_Bt}) multiplied by $L^2$ is equal to $iW_{3}(Z+i\epsilon)$. 
 
 As long as the $\text{AdS}_4$ background is stationary and stable we might use as well the ordinary Feynman diagrammatic technique in this case. Note that in the IR region we have
 \begin{equation}\label{asymptoic_of_wightman}
     W_{3}(Z) = \frac{1}{12\pi^2L^2}\frac{1}{Z^3} + \mathcal{O}\left(\frac{1}{Z^4}\right), \;Z\rightarrow \infty.
 \end{equation}
 Therefore, if there is actually a non-zero mass of the photon, we will find in its self-energy the term, proportional to the projector (\ref{projector_on_Bt}), and the following contribution in the effective action:
 \begin{gather}
     \delta\Gamma_{\text{eff}} = \frac{m^2_{\text{ph}}}{2}\int d\mu_{X}B_{A}\hat{\mathscr{P}}^{AB}_{t}B_{B} = \frac{m^2_{\text{ph}}}{2}\int d\mu_{X}\left[B^AB_{A} + X^2\partial^AB_{A} \frac{1}{ X^2\partial^2 - \left(X\cdot \partial\right)^2 - 3\left(X\cdot \partial\right)}  \partial^CB_{C}\right] \sim 
     \nonumber
     \\
     \sim \frac{m^2_{\text{ph}}}{2}\int d\mu_{X}d\mu_{Y} \partial^AB_{A}(X)\frac{i}{12\pi^2L^2} \frac{1}{Z^3} \partial^CB_{C}(Y),
\label{def_of_photon_mass}
 \end{gather}
 where in the last line we kept only the non-local part of its expression in the IR region. Now we straightforwardly integrate out the scalar fields with conformal mass in (\ref{bare_action_QED_AdS}) and omit the terms, proportional to $B^{A}B_{A}$ as they don't lead to the structure of the form (\ref{def_of_photon_mass}):
 \begin{gather}
     \delta\Gamma_{\text{eff}} = -i\frac{e^2}{2}\int d\mu_{X}d\mu_{Y}  B^A(X)B^{C}(Y)\frac{Y_{A}X_{C}}{L^4}\left[2W^{'}_cW^{*'}_c-W_cW^{*''}_c-W^{''}_cW^{*}_c\right] = \nonumber
     \\
     =  -i\frac{e^2}{2}\int d\mu_{X}d\mu_{Y}  \partial^AB_{A}(X)\partial^CB_{C}(Y)f(Z),
\label{effective_action_QED}
 \end{gather}
 where we have used that $Y_{A} = L^2\partial_{A}Z,\; X_C = L^2\partial_{C}Z$ and introduced the function $f(Z)$, which is the solution of $f^{''} = 2W^{'}_cW^{*'}_c-W_cW^{*''}_c-W^{''}_cW^{*}_c$, decaying at infinity. The solution of this differential equation indeed contains the term, proportional to $\frac{1}{Z^3}$: 
 \begin{equation}\label{f_asymptotic}
     f(Z) \simeq -\frac{1}{16\pi^4L^4}\frac{2}{3}\text{Re}\left(\alpha^{*}\beta\right)\frac{1}{Z^3}+\ldots\;\;.
 \end{equation}
 Substituting this result into the effective action (\ref{effective_action_QED}) and comparing the coefficients with (\ref{def_of_photon_mass}) we find
 \begin{equation}
     m^2_{\text{ph}} = \frac{e^2}{2\pi^2L^2} \text{Re}\left(\alpha^{*}\beta\right),
 \end{equation}
 which actually doesn't vanish when $\alpha$ and $\beta$ both are non-zero. 

 This result in AdS space at the simple example of the scalar QED confirms the discussions of the papers \cite{Porrati:2001db, Porrati:2003sa}, that such peculiarities of AdS as discrete spectrum of levels can lead to a presence of the Goldstone bosons as a bound states created by the electric current or stress-energy tensor even for free field  theory. Indeed, in the case of QED we see the propagator of the boson from $D(3,0)$ in the equations (\ref{def_of_photon_mass}), (\ref{effective_action_QED}), which is the analogy of the pole at $k^2 = 0$ in the Higgs mechanism in the Standard Model with the exchange of the massless field on fig.\ref{fig:Higgs}.
\bibliography{references}
\bibliographystyle{unsrt}
\end{document}